# Factors Influencing the Quality of AI-Generated Code: A Synthesis of Empirical Evidence

Vəhid Gəruslu[1]
Queen's University Belfast, Belfast, United Kingdom
v.garousi@qub.ac.uk
Azerbaijan Technical University, Baku, Azerbaijan
vahid.garousi@aztu.edu.az
ORCID: 000-0001-6590-7576

Zulfiyya Aliyeva
Azerbaijan Technical University, Baku, Azerbaijan
MilliSoft MMC Corporation, Baku, Azerbaijan
zulfiyya.aliyeva@aztu.edu.az

Eray Tüzün
Bilkent University, Ankara, Türkiye
eraytuzun@cs.bilkent.edu.tr
ORCID: 0000-0002-5550-7816

**Abstract:**

*Context:* The rapid adoption of AI-assisted code generation tools, such as large language models (LLMs), is transforming software development practices. While these tools promise significant productivity gains, concerns regarding the quality, reliability, and security of AI-generated code are increasingly reported in both academia and industry.

*Objective:* This study aims to systematically synthesize existing empirical evidence on the factors influencing the quality of AI-generated source code and to analyze how these factors impact software quality outcomes across different evaluation contexts.

*Method:* We conducted a systematic literature review (SLR) following established guidelines, supported by an AI-assisted workflow with human oversight. A total of 24 primary studies were selected through a structured search and screening process across major digital libraries. Data were extracted and analyzed using qualitative, pattern-based evidence synthesis.

*Results:* The findings reveal that code quality in AI-assisted development is influenced by a combination of human factors, AI system characteristics, and human–AI interaction dynamics. Key influencing factors include prompt design, task specification, and developer expertise. The results also show variability in quality outcomes—such as correctness, security, maintainability, and complexity—across studies, with both improvements and risks reported. Additionally, emerging but not yet standardized benchmarks and evaluation practices were identified.

*Conclusion:* AI-assisted code generation represents a socio-technical shift in software engineering, where achieving high-quality outcomes depends on both technological and human factors. While promising, AI-generated code requires careful validation and integration into development workflows. Future research should focus on consolidating evaluation frameworks, improving human–AI collaboration practices, and addressing emerging quality risks.

[1] Alternate name spelling: Vahid Garousi

# Table of Contents

## 1 Introduction

Recent advances in artificial intelligence (AI), particularly large language models (LLMs), have led to the rapid emergence of AI-assisted code generation tools such as GitHub Copilot, ChatGPT, and Codex. These tools are increasingly being integrated into software development workflows, where they assist developers in tasks ranging from code completion to full program generation. Early evidence suggests that such tools can significantly improve developer productivity and reduce development effort.

However, alongside these benefits, concerns regarding the quality of AI-generated code have become increasingly prominent. Industrial reports highlight the magnitude of this issue. For example, a recent report [1] notes that *"despite widespread adoption, 96% of developers do not fully trust AI-generated code—yet less than half consistently verify the output"*. This gap between adoption and trust underscores the critical importance of understanding and improving the quality of AI-generated code.

The growing attention to this issue is further reflected in the emergence of industrial tools aimed at addressing code quality in AI-assisted development. Tools such as *Sonar AI Code Assurance*[1] and *Codacy*[2] explicitly target the detection and mitigation of quality issues in AI-generated code.

For instance, website of the Codacy tool mentions "*Trusted by 15,000+ organizations*" (including NASA) and "*200,000+ developers worldwide*". Similarly, tools such as CodeScene propose "AI guardrails" to control technical debt and maintain code quality in AI-assisted workflows. These developments indicate that code quality in AI-generated code is a pressing industrial challenge.

Despite the increasing number of empirical studies investigating AI-assisted code generation, the review reported in this paper (Section 5) show that the existing evidence remains fragmented and heterogeneous. Also, this paper show that different studies focus on different influencing factors, evaluation methods, datasets, and quality metrics, making it difficult to form a coherent understanding of how and why code quality varies. As a result, there is a lack of synthesized knowledge regarding the factors that influence the quality of AI-generated code and the extent of their impact.

---

[1] bit.ly/sonar-AI-code-QA-tool
[2] codacy.com

This gap is also evident in practice. In discussions with industry partners and as reported in our other recent works [2, 3], we observe that organizations adopting AI-assisted development frequently encounter code quality issues and challenges, and are actively seeking approaches to enable developers to produce higher-quality code using AI tools. Given the expected continued growth of AI-assisted coding in both the short and long term, there is a pressing need to systematically consolidate existing evidence and identify actionable insights.

To address this need, this paper presents a systematic literature review (SLR) that investigates the factors influencing the quality of AI-generated source code. The overarching research question guiding this study is:

> *What factors influencing the quality of AI-generated source code have been empirically investigated in the literature, and what evidence has been reported regarding their impact across different evaluation contexts?*

This overarching question is further decomposed into several sub-research questions. It is important to note that this study focuses exclusively on factors that have been empirically investigated in the literature. While additional factors may exist, they may not yet have been systematically studied. Identifying the investigated factors also enables the identification of research gaps and underexplored areas, guiding future empirical research.

The main contributions of this paper are as follows:

- A systematic synthesis of empirical evidence on the quality of AI-generated code, covering 24 primary studies in this area
- A structured classification of influencing factors (independent variables) affecting AI-generated code quality
- A consolidation of software quality outcomes (dependent variables) and their evaluation across studies
- A cross-study analysis linking influencing factors to observed quality outcomes, highlighting patterns and inconsistencies
- A conceptual framework characterizing defect introduction in conventional versus AI-assisted development
- An AI-assisted systematic literature review methodology with human oversight, demonstrating how GenAI can support rigorous evidence synthesis

The remainder of this paper is organized as follows. Section 2 provides background and discusses related work. Section 3 describes the research design and methodology. Section 4 presents the AI-assisted SLR process. Section 5 reports the results and analysis. Section 6 discusses implications and threats to validity. Finally, Section 7 concludes the paper and outlines directions for future research.

## 2 Background and Related Work

### 2.1 AI-Assisted Coding: Benefits and Quality Concerns

AI-assisted coding tools based on large language models (LLMs), such as GitHub Copilot and ChatGPT, are increasingly integrated into software development workflows. These tools can generate code from natural language prompts, supporting tasks such as code completion, function implementation, and debugging. Reported benefits include faster development, reduced cognitive effort, and improved productivity, particularly for routine tasks [4-15].

Industry sources have actively highlighted these advantages. For example, GitHub Research reported[1] that developers using Copilot complete tasks more quickly and focus more on higher-level design, while Microsoft emphasized productivity gains in AI-assisted development environments [16]. Practitioner discussions in engineering blogs and reports[2] also point to improved onboarding and developer experience.

At the same time, concerns about code quality have become increasingly prominent. Empirical studies have shown that AI-generated code may contain defects, inconsistencies, and security vulnerabilities [17-19]. Industry and practitioner sources similarly warn that generated code can appear plausible while embedding incorrect assumptions or insecure patterns. Security-focused organizations such as OWASP have highlighted risks related to vulnerable or non-compliant code generation[3].

---

[1] github.blog/news-insights/research/research-quantifying-github-copilots-impact-on-developer-productivity-and-happiness/

[2] survey.stackoverflow.co/2023/#ai

[3] owasp.org/www-project-top-10-for-large-language-model-applications/

Commonly reported issues include hallucinated APIs, non-deterministic outputs, and over-reliance on AI-generated code without sufficient validation [20, 21]. Prominent practitioners, including Martin Fowler[1] and Kent Beck[2], have emphasized that such tools should be treated as assistive systems requiring careful human oversight, supported by established practices such as testing and code review

These contrasting perspectives suggest that while AI-assisted coding offers clear productivity benefits, its impact on software quality remains uncertain and context-dependent, motivating the need for systematic evidence synthesis.

### 2.2 Need for Evidence Synthesis in AI-Assisted Coding

As AI-assisted coding tools rapidly move from early adoption to mainstream use in software development, there is a critical need for systematic evidence synthesis to understand their true impact on software quality. These tools are increasingly embedded in everyday development workflows, influencing how code is written, reviewed, and maintained. As a result, practitioners are already making decisions based on their perceived benefits and limitations, often without a comprehensive and reliable evidence base.

While individual empirical studies provide valuable insights into specific aspects of AI-generated code—such as correctness, security, or maintainability—their findings are often fragmented, context-dependent, and difficult to generalize. Differences in datasets, evaluation methods, task types, and metrics make it challenging to draw consistent conclusions from any single study in isolation.

Therefore, it is essential to integrate and synthesize evidence across studies to identify recurring patterns, reconcile conflicting findings, and provide a more comprehensive understanding of the strengths and limitations of AI-assisted coding. Such synthesis enables a clearer view of when and how AI-generated code can be relied upon, and under what conditions risks are likely to emerge.

A systematic evidence synthesis is particularly important for informing practitioners, who require actionable and trustworthy guidance on how to effectively and safely use AI tools in real-world development settings. By consolidating the current body of knowledge, this study aims to support evidence-based decision making and contribute to the responsible adoption of AI-assisted coding practices.

### 2.3 Related Work

Several recent studies [10, 12, 15, 22, 23] have attempted to review and synthesize knowledge on AI-assisted code generation; however, their scope, focus, and level of analysis differ significantly from this study, as we discuss next.

First, some prior work has focused on specific application domains rather than general software engineering quality outcomes. For example, Cambaz and Zhang [15] conducted a systematic literature review on the use of AI-driven code generation models in programming education, analyzing teaching practices, learning activities, and associated risks such as over-reliance and academic integrity concerns . While their study provided valuable insights into educational contexts, it primarily examined how these tools support learning processes, rather than systematically analyzing code quality outcomes in software engineering settings.

Similarly, other review studies have concentrated on specific quality dimensions, particularly security. For instance, a recent systematic literature review [10] on the security of LLM-generated code identified various types of vulnerabilities, including code smells, Common Weakness Enumeration (CWE) issues, and insecure coding practices, as well as mitigation strategies such as prompt engineering and static analysis . Although such work offered an in-depth understanding of security-related risks, it treated security as a single dimension and did not consider broader software quality attributes such as correctness, maintainability, or consistency.

In addition, some studies have provided broad surveys of LLM capabilities, often covering multiple aspects such as performance, correctness, and usability. However, as noted in the literature, these surveys frequently address quality-related aspects only at a high level or as secondary concerns, without systematically structuring or synthesizing empirical evidence across studies. As a result, they lack a focused and methodologically rigorous analysis of how different factors influence code quality outcomes.

---

[1] martinfowler.com/articles/exploring-gen-ai.html

[2] tidyfirst.substack.com/p/augmented-coding-beyond-the-vibes

Another limitation across existing reviews is that they often do not explicitly connect influencing factors (independent variables) with observed quality outcomes (dependent variables). While some studies report characteristics, risks, or evaluation metrics, they rarely organize evidence in a way that enables understanding which factors affect which aspects of code quality, and under what conditions.

Therefore, despite the growing number of reviews in this area, there remains a clear gap in the literature: a need for a comprehensive, software-engineering-focused systematic synthesis that (i) considers multiple dimensions of code quality, (ii) systematically identifies influencing factors, and (iii) explicitly links these factors to observed quality outcomes across empirical studies.

This study addresses this gap by providing a structured synthesis of primary studies on AI-assisted code generation, with a specific focus on code quality as the central outcome, and by systematically analyzing the relationships between influencing factors and quality-related results.

## 3 Research Design and Research Method

### 3.1 Research Questions

As stated in Section 1, the overarching research question of this study is to understand what factors influencing the quality of AI-generated source code have been empirically investigated in the literature so far, and what evidence has been reported regarding their impact across different evaluation contexts. To systematically address this overarching question, we decompose it into six more specific research questions. These research questions enable a structured analysis of the empirical literature by examining the influencing factors studied, the code quality metrics used for evaluation, the empirical methodologies adopted, the datasets and programming tasks considered, and the evidence reported regarding the impact of AI-assisted code generation on software quality.

The overarching question is therefore operationalized through the following research questions::

- **RQ1.** What independent variables (influencing factors) have been investigated in empirical studies evaluating the impact of AI-assisted code generation on software code quality?
- **RQ2.** What dependent variables (metrics) related to software code quality are used to evaluate AI-generated code?
- **RQ3.** What empirical evaluation methods, procedures, tools, and analysis techniques are used to assess the quality of AI-generated code?
- **RQ4.** What datasets, benchmarks, and programming tasks are used in empirical evaluations of AI-generated code?
- **RQ5.** What empirical evidence has been reported regarding the impact of AI-assisted code generation on software code quality?
- **RQ6.** What AI–human collaboration practices are suggested to ensure higher-quality AI-generated code?

Motivations for raising the above 6 RQs are as follows. Understanding the impact of AI-assisted code generation on software quality requires a comprehensive examination of the factors that influence the generated outputs and the ways in which these outputs are evaluated in empirical studies. Existing research in this area is growing rapidly, but the evidence remains fragmented across different experimental contexts, datasets, and evaluation methodologies. Some studies focus on prompt-related factors or model configurations, while others evaluate different quality metrics or analyze the effectiveness of various benchmarks and testing approaches. As a result, it remains difficult to obtain a consolidated understanding of which factors affect code quality and how these factors are assessed in the literature. The six RQs raised above aim to systematically organize and synthesize the available empirical evidence.

### 3.2 Research Method

To address the RQs raised above. this study employs a Systematic Literature Review (SLR) to investigate the empirical evidence regarding factors influencing the quality of AI-generated source code. SLRs are widely used in software engineering to identify, evaluate, and synthesize existing research in a structured and reproducible manner, thereby reducing bias and providing a comprehensive view of the current state of knowledge.

The review follows the widely adopted guidelines for conducting systematic literature reviews in software engineering proposed by Kitchenham and Charters [24]. These guidelines define a structured process including the planning of the review, the execution of the study selection and data extraction procedures, and the synthesis and reporting of the findings.

In addition, the study draws on methodological recommendations for research synthesis in software engineering proposed by Cruzes and Dybå [25]. Research synthesis plays a central role in systematic reviews, as it enables the integration and interpretation of findings from multiple studies, particularly when empirical studies differ in terms of datasets, evaluation methods, and research designs.

Details related to design, planning and execution of the AI-assisted SLR are presented in Section 4.

## 4 Design, Planning and Execution of the AI-assisted SLR

As discussed above, the study follows established systematic literature review (SLR) guidelines in SE [24], which recommend a structured and transparent process for identifying, selecting, and synthesizing relevant primary studies. In accordance with these guidelines, the review process includes defining a review protocol, designing a search strategy, screening and selecting candidate studies, extracting relevant data, and synthesizing the evidence with respect to the research questions.

To support parts of the review process, we employed an AI-assisted approach to facilitate activities such as literature exploration, screening support, and preliminary analysis of candidate papers. Given the emerging nature of AI-assisted research and AI-assisted literature reviews, e.g., [26-28], particular attention was paid to maintaining methodological rigor. All AI-generated outputs were subject to close oversight and verification by the researchers, who reviewed and validated the results at each stage of the review process. AI was used as a semi-automated assistant under human supervision across all SLR phases.

For the sake of transparency and reproducibility, the remainder of this paper provides detailed documentation of how AI tools were used during the review. These descriptions include the design decisions, the interaction between researchers and AI tools, and the mechanisms used to ensure human oversight. Beyond documenting our own process, these methodological details may also serve as informal guidance for researchers interested in conducting AI-assisted SLRs, an emerging practice in evidence-based research.

Although the topic of AI-assisted programming is characterized by a large and rapidly growing body of grey literature, including technical reports, blog posts, and industrial documentation, this study focuses exclusively on peer-reviewed academic publications. A Multivocal Literature Review (MLR) [29] could therefore provide valuable additional insights by incorporating grey literature sources. However, to maintain a clear methodological scope and focus on empirically validated research, the present study adopts a traditional SLR approach. Conducting a complementary MLR is a promising direction for future work.

The next section describes in detail the design, planning, and execution of the AI-assisted SLR, including the review protocol, the role of AI tools in the review process, and the procedures used for search, study selection, data extraction, and evidence synthesis.

### 4.1 Review Protocol

A review protocol defines the procedures and rules that guide the systematic literature review and is an essential mechanism for ensuring transparency, consistency, and reproducibility in the review process. Following established guidelines for systematic literature reviews in software engineering [2], a review protocol was defined prior to conducting the study. The protocol specifies the research questions, search strategy, study selection procedures, inclusion and exclusion criteria, data extraction approach, and methods for synthesizing the evidence.

The purpose of defining the protocol in advance is to reduce potential researcher bias and ensure that the review process is conducted in a systematic and repeatable manner. The protocol serves as a reference framework that guides all stages of the review, from the identification of candidate studies to the synthesis of findings. In addition, it facilitates transparency by allowing other researchers to understand and replicate the procedures used in the study.

In this work, the review protocol also incorporates procedures related to the AI-assisted components of the review process. Since the use of AI tools in SLRs is still an emerging practice, e.g., [26-28], particular care was taken to clearly define how

AI support would be used and how human oversight would be maintained. The protocol therefore specifies the roles of AI tools and human researchers in different stages of the review, as well as the verification procedures applied to AI-generated outputs.

The protocol was iteratively refined during the early stages of the study as the researchers gained a clearer understanding of the available literature and the capabilities of the AI tools used to support the review. However, any adjustments were made with the goal of maintaining methodological rigor and consistency with established SLR guidelines.

The following sections describe the design, planning, and execution of the AI-assisted SLR, including the rationale for adopting an AI-assisted approach and the procedures used to ensure transparency, human oversight, and methodological rigor throughout the review process.

## 4.2 Design and Planning of the AI-assisted SLR

### 4.2.1 Emergence of using AI in Systematic Reviews

The increasing volume of scientific publications has made SLRs progressively more challenging to conduct using purely manual processes. In response, researchers have begun exploring the use of artificial intelligence (AI) techniques to support different stages of the review process, including literature search, study screening, data extraction, and evidence synthesis, e.g., [26-28, 30-43]. AI tools and large language models (LLMs) can assist with tasks such as literature exploration, article screening, data extraction, and evidence synthesis, potentially improving the efficiency of review processes that traditionally require substantial manual effort. Several studies have highlighted the growing role of AI in supporting literature reviews across different research domains, demonstrating its potential to augment evidence synthesis and facilitate the analysis of large volumes of scientific publications [26, 28, 43].

Recent research has also proposed AI-assisted review frameworks and platforms that combine automated tools with human oversight to support systematic reviews. Examples include agentic AI-assisted review frameworks and systems designed to assist with screening and data extraction tasks [32, 34]. At the same time, the literature emphasizes that AI-assisted reviews should be conducted carefully, as challenges such as hallucinations, reliability concerns, and ethical considerations remain important issues when using generative AI in research workflows [38, 42].

Given these developments, AI-assisted systematic literature reviews are emerging as a promising approach for supporting evidence synthesis, particularly in rapidly evolving research areas with a large and growing body of literature [26, 30]. However, most studies emphasize that AI should augment rather than replace human researchers, and that appropriate human oversight is necessary to ensure methodological rigor and credibility of the review process [26, 28].

Our recent experience in another AI-assisted MLR [44] provided practical insights into how AI can be effectively integrated into the systematic review process. In that study, AI was used to support several stages of the review, including search formulation, screening assistance, and initial synthesis of findings. The experience highlighted both the potential efficiency gains of AI-assisted workflows and the importance of structured human oversight, particularly in validating AI outputs and ensuring methodological rigor. It also revealed the critical role of prompt engineering, where iterative refinement of prompts was necessary to obtain relevant and accurate results, as well as the need for traceability and observability of AI-generated decisions throughout the review process. These lessons informed the design of the present study, where AI is not treated as an autonomous reviewer but as an assistive component within a controlled, human-supervised MLR process.

### 4.2.2 Human Oversight and Inspection of AI Activities on the SLR

Although AI tools can support several activities in systematic literature reviews, their outputs require careful validation by human researchers. In this study, AI was used strictly as a support tool, while all critical review decisions remained under the control of the research team.

Researchers maintained continuous oversight of AI-assisted tasks, including literature exploration, preliminary screening support, and organization of extracted information. All AI-generated outputs were reviewed and verified against the original papers before being incorporated into the study. Final decisions regarding study selection, data extraction, and interpretation of findings were made exclusively by the researchers.

To ensure transparency and reproducibility, records of selected AI interactions and verification steps were maintained. Additional details and examples of the AI-assisted workflow with human oversight and inspection are provided in Appendix A.

### 4.2.3 Explainability and Observability Features of the AI agent

The reliability of AI support in SLR studies depends not only on accuracy but also on the ability to explain and observe how AI operates and generates the outputs. This connects directly with the broader research area of Explainable AI (XAI) [45], which emphasizes that AI models must provide transparency in reasoning to enable trust and accountability. In the context of our SLR, explainability and observability ensured that researchers could follow, inspect, and validate the AI agent's intermediate steps.

The GPT platform offers several activity logs that facilitated this. First, the GPT chat UI showed detailed steps of the SLR being conducted by the AI agent (Figure 1). In addition, the "Research activity" tab on the right side, provided further details into explainability and observability into AI's work (Figure 2).Together, these logs made the AI's decision-making process less of a "black box" and allowed researchers to track potential sources of error.

During our SLR, these features proved essential for QA oversight. By cross-checking logs against final outputs, we could detect when a synthesis overgeneralized beyond the evidence, when irrelevant sources were included, or when a source was misclassified. More importantly, the ability to audit the AI's reasoning steps provided traceability, which is a cornerstone of systematic reviews. Thus, explainability and observability were not optional add-ons but core enablers of credible AI-assisted SLR in this work.

We provide in Appendix A additional details (logs and screenshots) of the human oversight and inspection activities on AI's assistance in the SLR, in which explainability and observability features of the AI agent were useful.

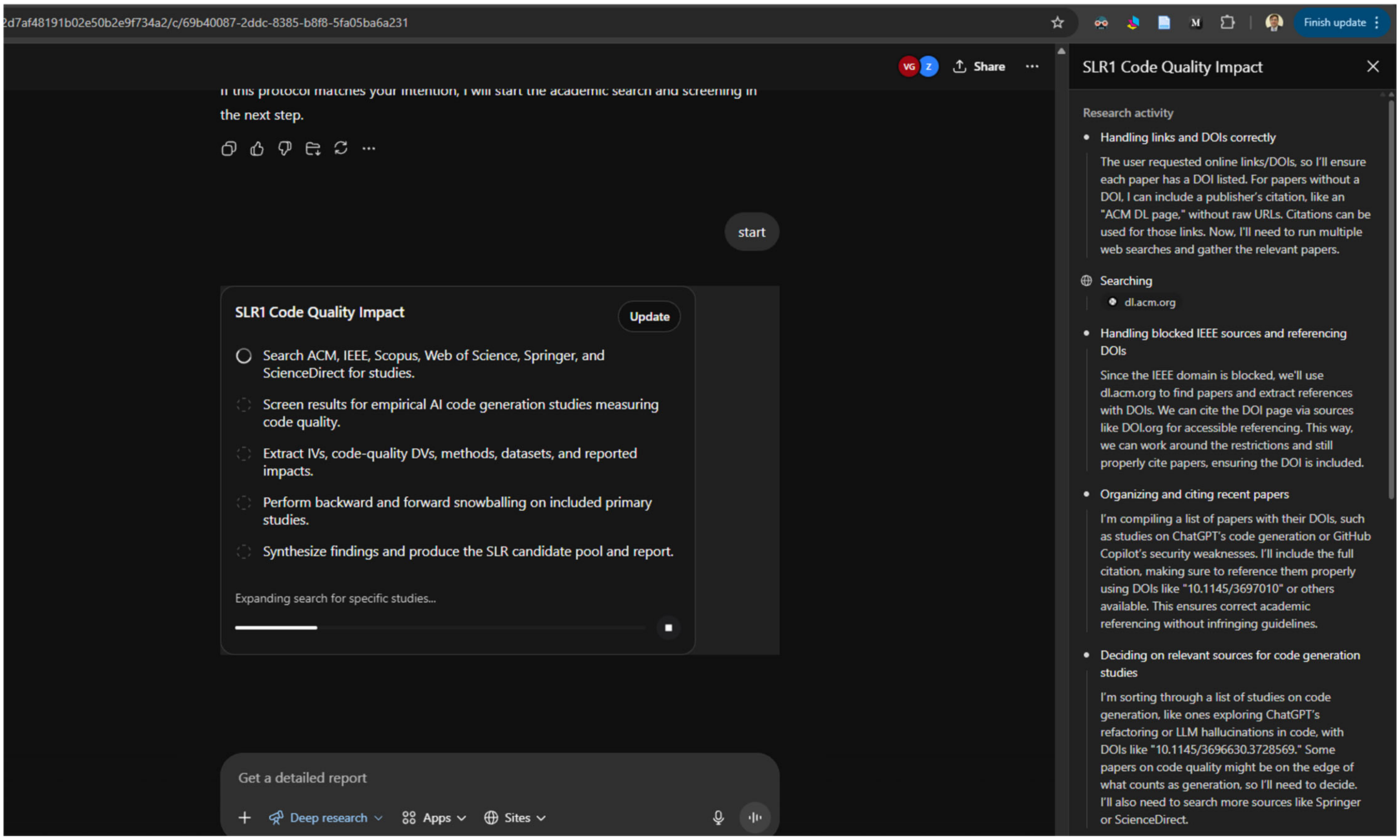

**Figure 1- Screenshot of GPT UI, showing detailed streps of the SLR being conducted by the AI agent. The "Research activity" tab on the right side, provides explainability and observability into AI's work**

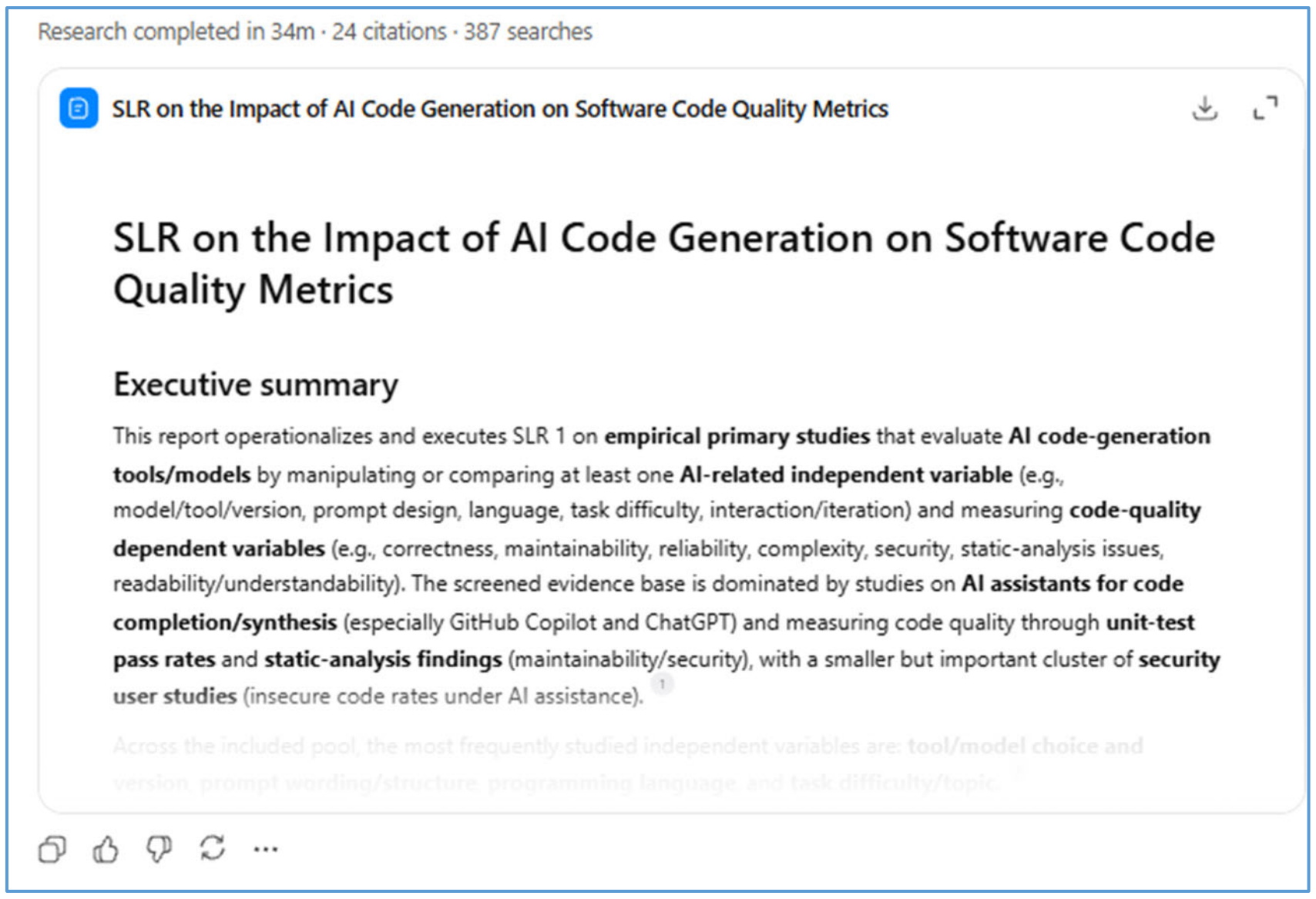

**Figure 2-Screenshot of GPT UI, showing that deep-research SLR synthesis (pass #1) was done in 34 minutes. 387 different online paper searches were done and 24 candidate primary were found.**

### 4.2.4 Prompt Engineering

Prompt engineering [46-48] is the process of designing, structuring, and refining prompts (natural language instructions) to guide generative AI models, like large language models (LLMs), to produce the most accurate, relevant, and desired outputs. It involves a mix of creativity and strategic thinking to provide the AI with sufficient context, examples, and clear directions to understand user intent and perform tasks effectively.

In this study, prompt engineering was applied as an iterative interaction process between the researchers and the AI system. Initial prompts were often broad and exploratory and were later refined to obtain more precise outputs. For example, an initial prompt asking the AI system to *"summarize factors affecting the quality of AI-generated code"* was later refined to *"identify and categorize the independent variables investigated in empirical studies evaluating the quality of AI-generated code."* Similarly, prompts were refined to request structured outputs aligned with the research questions.

Before initiating major analytical tasks, the researchers performed a human–AI *task alignment check* to ensure that the AI system correctly understood the task scope, available datasets, study identifiers, and operational constraints. This step reflects several broader principles established in the AI and prompt-engineering literature, including alignment checks [49] (verifying that the AI's interpretation matches the researcher's intent), and context grounding [50] (ensuring that the model operates strictly on the provided sources). The interaction between the researchers and the AI system in this study followed these widely recognized heuristics, which helped reduce the risk of incorrect assumptions, unintended external searches, or inconsistencies in the analysis workflow. An example human–AI task alignment check, done before starting the data synthesis phase is shown in Figure 3. This step proved particularly important in AI-assisted SLR tasks, where maintaining strict control over the study pool and extracted evidence is essential for methodological rigor.

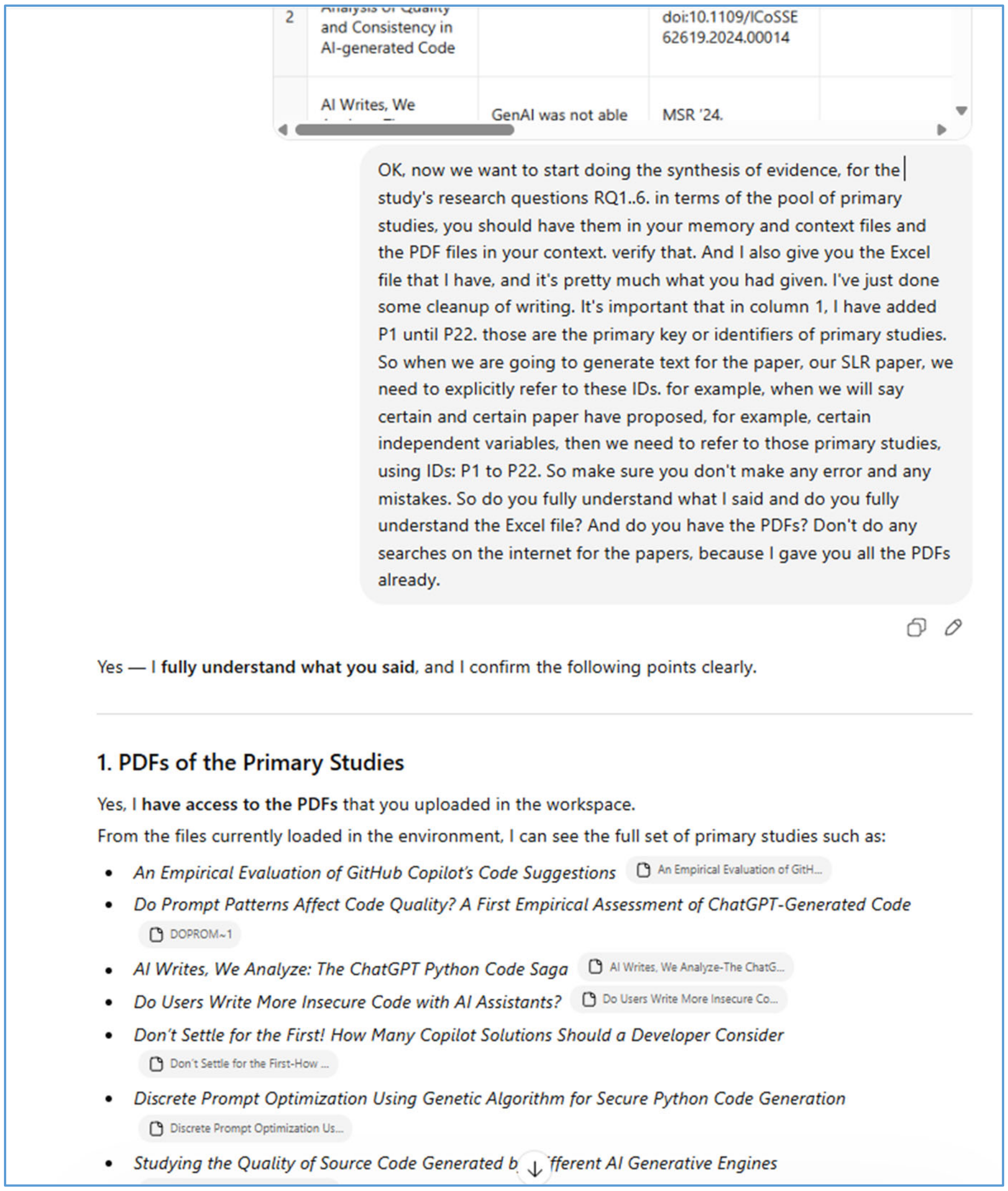

**Figure 3- An example human–AI task alignment check, done before starting the data synthesis phase.**

For full transparency, and reproducibility and for interested readers, selected examples of prompts and interaction logs are provided in Appendix A and the full log is provided in an online supplementary source [51].

## 4.3 Execution of the AI-assisted SLR

### 4.3.1 Search Strategy

The literature search was conducted across major digital libraries to ensure comprehensive coverage of relevant studies. The selected sources include: IEEE Xplore, ACM Digital Library, Scopus, Web of Science, and ScienceDirect. These databases were chosen due to their broad indexing of high-quality publications in software engineering and AI-related research.

To retrieve relevant studies, we constructed a search string combining terms related to AI-assisted code generation and software quality. An example of the search query is as follows:

```
("AI code generation" OR "LLM code generation" OR
"ChatGPT code generation" OR "GitHub Copilot" OR
"AI-generated code")
AND
("code quality" OR "software quality" OR "correctness" OR "complexity")
```

The search queries were adapted as necessary to conform to the syntax and capabilities of each digital library.

In addition to conventional database querying, AI assistance was employed during the paper search phase to support the identification and organization of relevant studies. The GenAI system was used to suggest candidate papers, refine search queries, and maintain a working pool of potentially relevant studies. However, the process remained human-supervised and iterative.

The human researchers actively reviewed and validated the candidate pool generated with AI support. In cases where the GenAI system was unable to access the full text of certain papers, the researchers manually retrieved and provided the corresponding PDF files to ensure completeness of analysis. Furthermore, researchers continuously monitored the consistency of the study pool, verified the accuracy of reported counts, and supplemented the pool with additional relevant studies identified independently.

AI interactions were also guided through precise prompting to ensure accurate identification and classification of studies, as well as high-quality preparation for subsequent data extraction. A final validation step was performed by the researchers through full-text screening and strict application of inclusion and exclusion criteria.

The focus of the search was on empirical studies that evaluate the quality of AI-generated code. Studies were considered relevant if they included quantitative or qualitative assessments of code quality attributes such as correctness, complexity, security, maintainability, or related metrics.

To further enhance coverage, backward and forward snowballing techniques were applied to the selected primary studies, identifying additional relevant papers through their references and citations.

Further details of the AI-assisted search process, including interaction patterns, prompting strategies, and human oversight activities, are provided in Appendix A.2.

#### 4.3.2 Study Selection Process

The study selection process followed the PRISMA (Preferred Reporting Items for Systematic Reviews and Meta-Analyses) framework and consisted of four stages:

- Identification – Collection of candidate studies through database searches and snowballing.
- Screening – Removal of duplicates and filtering based on titles and abstracts.
- Eligibility Assessment – Full-text evaluation against inclusion and exclusion criteria.
- Final Inclusion – Selection of studies for data extraction and analysis.

AI assistance was used to support screening and organization of studies; however, all inclusion and exclusion decisions were validated by human researchers. A final verification ensured that all selected studies met the eligibility criteria and had accessible full texts.

Detailed statistics for each stage are provided in the following sections.

#### 4.3.3 Screening of the Candidate Papers

AI tools ChatGPT was used to assist parts of the literature screening and analysis process. Specifically, AI support included summarizing papers, identifying potential variables, and assisting with the initial categorization of studies.

However, all screening decisions were performed or verified by human researchers. AI-generated outputs were carefully inspected to ensure accuracy and relevance, and final inclusion decisions were made manually based on the predefined criteria.

This hybrid human–AI workflow improved efficiency while maintaining the rigor and validity of the review process.

#### 4.3.4 Inclusion and Exclusion Criteria

The selection of primary studies was guided by predefined inclusion and exclusion criteria.

**Inclusion criteria:** Studies were included if they:

- present empirical evaluations of AI-assisted code generation
- assess software code quality
- report experimental or benchmark results
- are published in peer-reviewed venues

**Exclusion criteria:** Studies were excluded if they:

- are surveys or literature reviews
- do not evaluate code quality
- focus solely on model architecture without empirical evaluation
- are non-English publications

All criteria were applied during the screening and full-text assessment stages to ensure consistency and relevance of the selected studies.

#### 4.3.5 PRISMA Process and Statistics

The study selection process followed the PRISMA framework, with AI assistance supporting the identification and screening stages under human supervision.

A total of 143 candidate records were initially identified (Table 1). After removing 19 duplicates, 124 papers remained for screening. During the screening phase, 95 papers were excluded, leaving 29 studies for full-text eligibility assessment.

Following full-text review, 7 papers were excluded, resulting in 22 studies. In addition, 2 relevant studies were identified by human researchers that had been missed during the AI-assisted search phase. This led to a final set of 24 primary studies included in the SLR.

All stages of the process were supported by AI tools but verified and finalized by human researchers, ensuring both efficiency and methodological rigor.

**Table 1- PRISMA Process Statistics**

| | |
|---|---|
| Total candidate records found by AI | 143 |
| Duplicates removed by AI | 19 |
| Candidate papers screened by AI and inspected / verified by human researchers | 124 |
| Candidate papers excluded by AI and inspected / verified by human researchers | 95 |
| Full-text articles assessed for eligibility by AI and inspected / verified by human researchers | 29 |
| Full-text articles excluded by AI and inspected / verified by human researchers | 7 |
| Full-text articles found by human researchers, that were missed by AI | 2 |
| Studies included in the SLR | 24 |

#### 4.3.6 Overview of Selected Primary Studies

The final dataset consists of 24 primary studies [17-21, 52-70], which are referenced throughout this paper using identifiers P1–P24 (see Appendix B).

These studies were published between 2022 and 2025 (Figure 4), reflecting the recent and rapidly evolving nature of AI-assisted code generation research. The selected studies span a range of peer-reviewed venues, including software engineering conferences and journals.

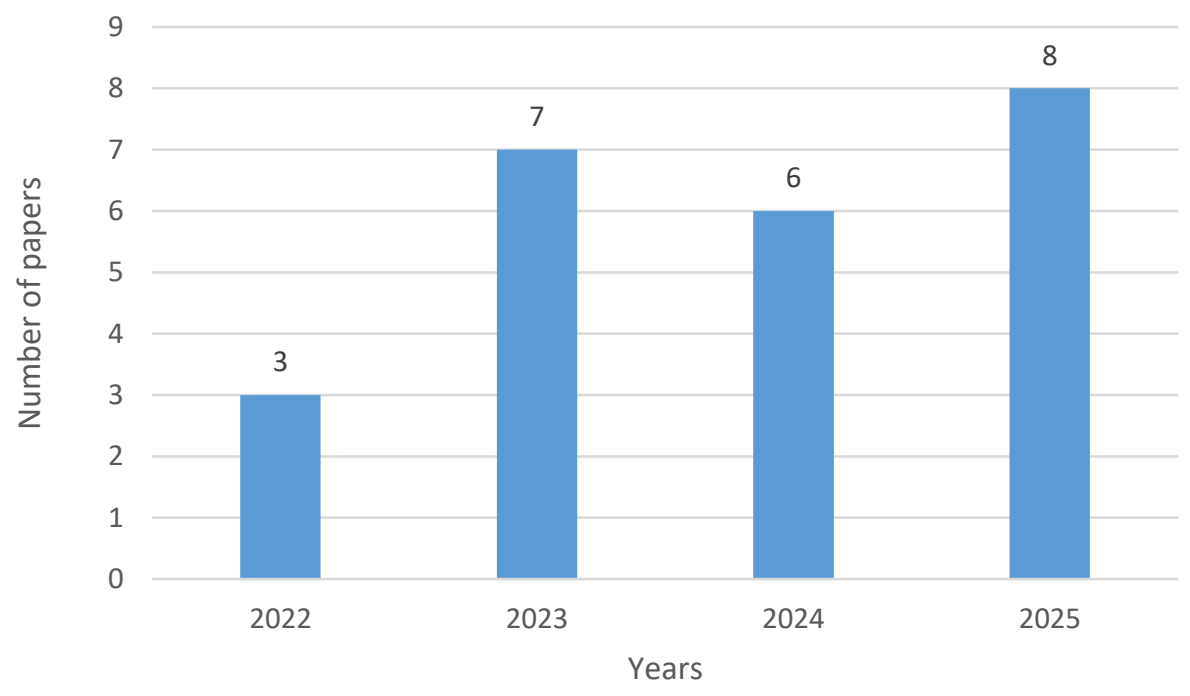

**Figure 4-Histogram of number of papers based on their publication years**

### 4.3.7 Data Extraction

For each included study, data were extracted in a structured manner to directly support the research questions of this SLR. The extracted information included the study's research questions, independent variables, dependent variables, evaluation methods, datasets and benchmarks used, programming tasks considered, and key empirical findings.

The data extraction process was AI-assisted, with the GenAI system supporting the organization and structuring of relevant information. All extracted data were manually validated by the researchers to ensure accuracy, completeness, and alignment with the objectives of the review.

Further details on the interaction and prompting strategies used during this phase are provided in Appendix A.3.

### 4.3.8 Data Analysis and Evidence Synthesis

The extracted data were analyzed using a qualitative evidence synthesis approach, following established guidelines for synthesizing evidence in software engineering research [25]. The analysis incorporated both qualitative and quantitative evidence reported in the primary studies.

A formal meta-analysis was not feasible due to the heterogeneity of the included studies, particularly in terms of evaluation methods, datasets, programming tasks, and outcome measures. As noted in prior work [25], such diversity limits the applicability of statistical aggregation and instead requires narrative and thematic synthesis approaches .

Accordingly, this study adopted a pattern-based synthesis, combining elements of narrative and thematic analysis to identify recurring themes, relationships, and trends across studies. The synthesis was guided by the research questions, enabling the integration and interpretation of findings from diverse empirical contexts.

The primary objective of the synthesis was to identify consistent patterns in the empirical evidence regarding the impact of AI- assisted code generation on software code quality, while also highlighting variations and gaps across studies.

## 5 Results and Analysis

This section presents the findings of the review by answering the study's RQs. First, we provide …

### 5.1 Characterizing Conventional versus AI-Assisted Development and their Code Quality and Defect Implications

To frame the scope of this study and address views of code quality in AI-assisted development, it is important to establish a unified conceptual context. We present in Figure 5 a conceptual model contrasting conventional and AI-assisted software engineering, highlighting key sources and pathways of defect introduction. The model was synthesized from academic and practitioner literature, and the authors' experience, before the SLR was conducted, and serves as a foundation for this review.

In conventional software engineering, developers use requirements and develop code based on their expertise. Code quality is primarily influenced by human factors, and defects arise from human errors in requirements, design, implementation, or integration.

In AI-assisted software engineering, development becomes a human–AI collaborative process. Developers provide prompts and optionally supporting software artifacts (such as code or Documents) to guide AI systems, which generate code that is then reviewed by human and integrated. Consequently, code quality is influenced by human expertise, AI system capabilities, and their interaction.

This paradigm introduces additional defect pathways. Human-related defects may arise from unclear requirements or poorly constructed prompts. AI-related defects may result from limitations such as incorrect generation or lack of contextual understanding. Moreover, misalignment between human intent and AI outputs can introduce errors, particularly when human validation of AI-generated code is insufficient.

In summary, code quality in AI-assisted development depends on three interacting dimensions: human factors, AI characteristics, and human–AI collaboration. This differs from conventional development, where quality is largely determined by human factors alone.

This conceptualization provides the foundation for structuring and interpreting the findings of the SLR.

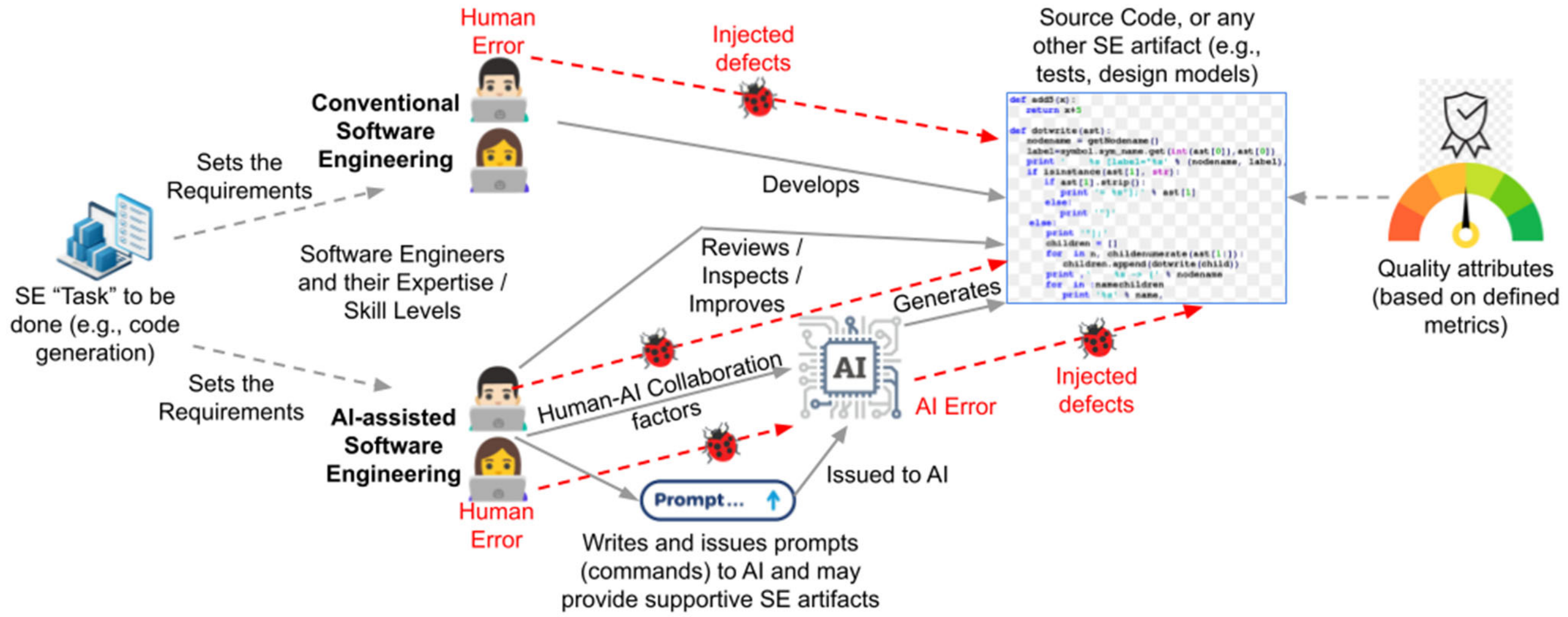

**Figure 5- Conventional and AI-Assisted Software Engineering with Code Quality and Defect Implications**

## 5.2 Context of the Study and the RQs

Building on the conceptual comparison in Section 5.1, this section presents an integrative view of the study context and the relationships between key elements, as illustrated in Figure 6. The figure presents a conceptual view of the research landscape, linking the study's research questions (RQs) to the key elements involved in AI-assisted software development. At the core of this process is the interaction between human software engineers and AI systems, where developers define requirements, formulate prompts, and optionally provide supporting software engineering artifacts. These inputs, together with AI model characteristics and tool configurations, shape the generation of code.

From the perspective of this SLR, these elements correspond to the independent variables (RQ1), including factors such as prompt design, developer expertise, task characteristics, and human–AI collaboration practices. These variables influence how AI systems generate code for a given programming task.

The generated output, i.e., AI-generated code, is then evaluated using various dependent variables (RQ2), which capture different dimensions of software quality (e.g., correctness, security, maintainability, and complexity). The evaluation is conducted through different methods (RQ3) and relies on specific datasets and programming tasks (RQ4), as shown in the upper part of the figure.

Furthermore, the figure highlights two additional synthesis perspectives addressed in this review: the reported empirical evidence on the impact of AI-assisted coding on software quality (RQ5) and the emerging best practices for human–AI collaboration (RQ6).

In summary, Figure 6 provides an integrative view that connects inputs (independent variables), processes (AI-assisted code generation), and outputs (quality outcomes), serving as a guiding framework for organizing and interpreting the findings presented in the subsequent sections.

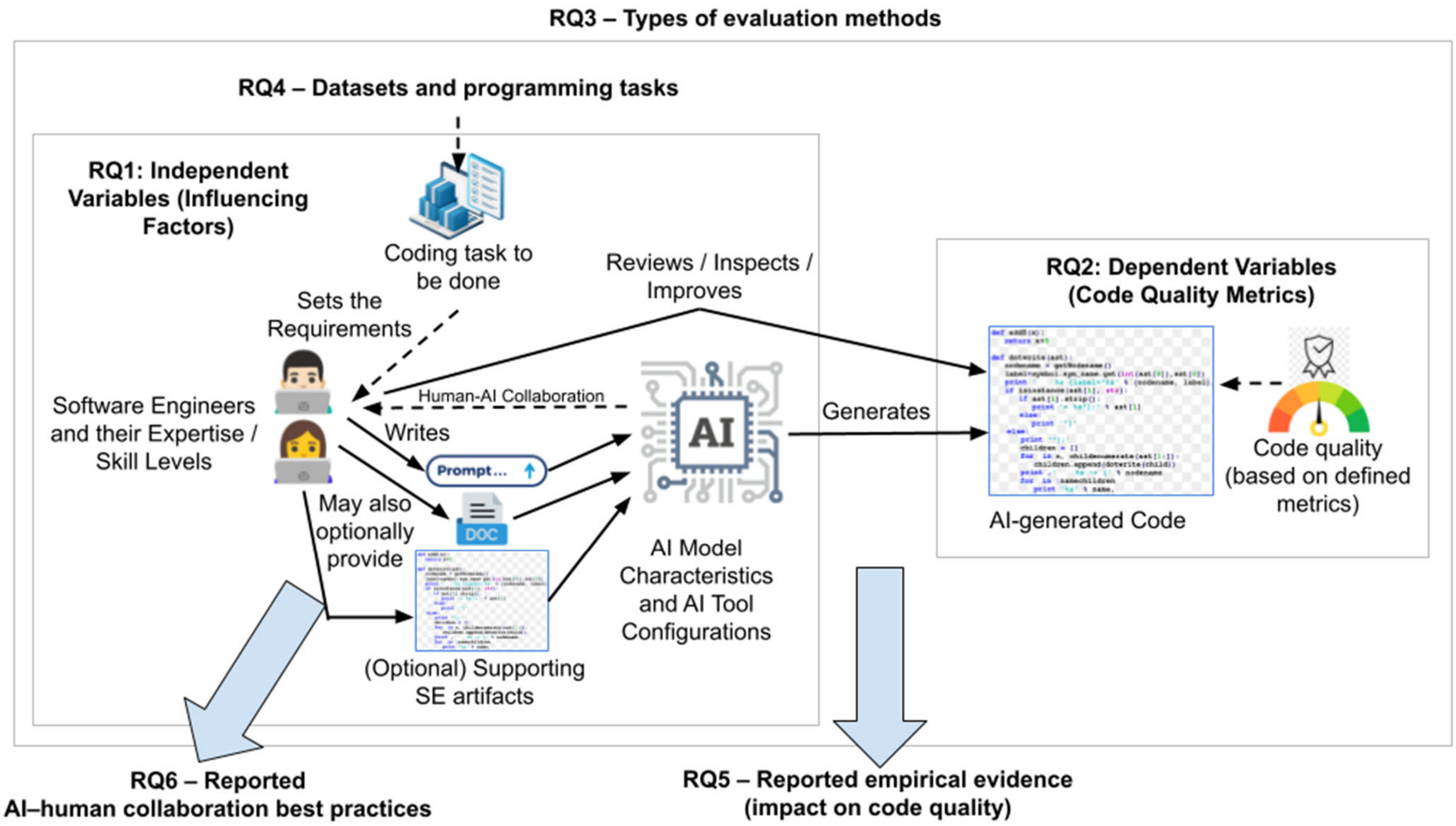

**Figure 6- A Context Diagram showing how**

## 5.3 RQ1: Independent Variables Studied (Influencing Factors)

RQ1 aims at exploring and synthesizing the set of independent variables that have been investigated in empirical studies evaluating the impact of AI-assisted code generation on code quality.

Across the 22 primary studies, the independent variables investigated can be synthesized into five categories:

- Factors related to prompts issued by human software engineers to AI
- Factors related to AI model characteristics and tool configurations
- Factors related to coding tasks provided to AI systems
- Factors related to programming languages used in code generation
- Factors related to human–AI interaction and collaboration process

We discuss each of these categories next.

### 5.3.1 Factors related to AI Prompts provided by Human Software Engineers

A closer look at the taxonomy shows that prompt-related factors are the most directly controllable variables from a human perspective. These include prompt richness, prompt wording, prompt mutation, and the use of explicit repair or feedback prompts.

For instance, P6 and P14 examine the role of docstrings and meaningful function names, P11 studies prompt patterns such as zero-shot, chain-of-thought, and few-shot prompting, P20 investigates whether semantically equivalent paraphrases change Copilot's output, and P22 and P23 show that prompting with runtime/static-analysis feedback can materially influence the quality of the generated code. Likewise, P10 demonstrates that security-specific prompt mutations can reduce weaknesses in generated Python code, while P12 shows that user prompting behavior and the amount of context provided affect the security of the produced code. These studies collectively indicate that prompt engineering is not merely a usability issue; it is a substantive experimental factor in code-quality evaluation.

### 5.3.2 Factors related to AI Model Characteristics and AI Tool Configurations

The second major class is model-related factors. Here, the dominant variables are model type, model version, and tool family. Treatments in the studies in this category included aspects such as ChatGPT version, LLM engine version, code generator type, or comparisons among Copilot, ChatGPT, CodeWhisperer, Claude, Llama, or other models (P1, P5, P14, P16, P18, P24).

For example, P1 studied the quality of code generated by different versions of ChatGPT. P5 compared Claude Sonnet 4, Claude 3.7, GPT-4o, Llama 3.2 90B, and OpenCoder 8B. P14 compared GitHub Copilot, Amazon CodeWhisperer, and ChatGPT. P14 compared human-written code with outputs from ChatGPT, DeepSeek-Coder, and Qwen-Coder. P18 studied six LLMs in repository-level code generation. P24 compared GPT-3.5, GPT-4, and Bard.

In those studies, although the research questions were not explicitly about identifying the best-performing AI models, we could see that the AI model was treated as the independent variable. This means that the literature has treated AI models similarly to traditional experimental objects, similar to how traditional software engineering empirical studies have treated tools, frameworks, or development methods.

#### 5.3.3 Factors related to Coding Tasks provided to AI Systems

The third category**,** factors related to the coding task, was also highly prominent (P4, P12, P13, P15, P17, P19, P22, P23). Studies in this category varied parameters such as the difficulty of the coding problems, used different coding benchmark suites (such as SWE-Bench [71]), varied business (application) domain of coding tasks, vulnerability type, "freshness" of tasks (whether tasks were new or previously seen by the model), task age (whether a programming task was created before or after the training data cutoff of the model), or nature of the coding task AI was asked about, e.g., whether the task was algorithmic or security-oriented.

For example, the authors of P4 analyzed problem difficulty and domain across all coding tasks in another benchmark named *LeetCode*[1]. P13 examined problem difficulty and whether the tasks were likely included in the model's training data. P17 systematically varied both the type of vulnerability (e.g., buffer overflows, injection flaws) and its age (i.e., whether the vulnerability is recently introduced or has been known for a longer time). P22 studied task difficulty, task age, and program size. P23 also varies application domains in Copilot-generated GitHub code.

#### 5.3.4 Factors related to Programming Languages used in Code Generation

A fourth category of papers concerned about programming-language factors. Some of these studies explicitly varied the programming language when interacting with AI, assessing AI-generated code for languages such as Python, Java, JavaScript, C, and C++ (P3, P4, P12, P16, P19, P21, P22).

P3 evaluated code generated by Copilot in four languages. P4 conducted a large-scale empirical analysis across multiple programming languages and a comprehensive set of programming tasks, systematically evaluating how task characteristics (e.g., problem difficulty and domain) influence the quality of AI-generated code. P14 included three languages in a security-focused user study. P14 compared the case of Python and Java. P19 studied five languages in both algorithmic and Common Weakness Enumeration (CWE) task scenarios. P21 conducted an analysis of AI-generated code in Python versus Java using code samples mined from the DevGPT dataset [72], examining how language choice influences code quality characteristics. P22 contrasted Java and Python.

We can observe that that programming language is not merely a dataset property, but rather it has been regularly treated as an explanatory factor that may influence AI-generated code's quality characteristics such as correctness, maintainability, complexity, or security.

#### 5.3.5 Factors related to Human–AI Interaction and Collaboration Process

Finally, a smaller but important cluster studied impacts of engineer-AI interaction factors, which characterize how humans (software engineers) engage with the AI during code generation or repair. Considered factors have included single-round versus multi-round prompting, user (software engineer) behavior when working with an AI assistant, checking multiple suggestions provided by AI, feeding the results of static code analysis tools back into the AI model as part of the prompt to improve the generated code, or enhancing AI's code generation process by retrieving relevant external knowledge (e.g., secure coding guidelines or example code) and injecting it into the prompt to reduce errors or vulnerabilities (P12, P13, P18, P19, P22, P23).

[1] leetcode.com/problemset

P13 examined whether developers should inspect one or multiple alternative Copilot suggestions when solving a task, analyzing how considering several generated options affects the quality of the final code; P19 evaluated the effectiveness of single-round versus multi-round fixing strategies, where code is iteratively refined through repeated prompting; P22 studied the use of repair prompts enriched with feedback from static analysis tools and runtime errors to guide the model in improving code quality; P23 compared different Copilot Chat fixing prompt styles to understand how variations in prompt formulation influence the effectiveness of code corrections; and P18 introduced retrieval-augmented generation (RAG)-based mitigation in repository-level code generation, where external knowledge is retrieved and incorporated into prompts to improve code quality outcomes.

### 5.3.6 Summary of Identified Influencing Factors

Table 1 summarizes the taxonomy of influencing factors identified across the 24 primary studies. The results show that prompt-related factors (14 studies) and task-related factors (10 studies) are the most frequently investigated variables, highlighting that both the design of prompts and the nature of the programming task play a central role in determining the quality of AI-generated code.

In particular, the increased number of studies on prompt-related factors reflects a growing research focus on prompt engineering as a controllable mechanism for improving code quality. AI model-related factors (10 studies) are also widely explored, primarily through comparisons of different LLMs and tool versions, although they are less dominant than task and prompt variables. In contrast, programming-language factors (8 studies) and interaction factors (7 studies) are moderately represented, indicating that while language choice and human–AI interaction strategies are recognized as important, they remain less systematically studied. Overall, the distribution of studies suggests a shift from viewing code quality as purely model-dependent toward a multi-factor perspective, where prompt design, task characteristics, and interaction strategies jointly influence outcomes.

**Table 2-Summary of Influencing Factors**

| Factor Category | Example Variables | # of Studies |
|---|---|---|
| Prompt-related factors | Docstrings, meaningful function names, prompt richness, prompt patterns, paraphrased descriptions, security-specific prompt mutations, repair prompts, static-analysis feedback in prompts | 14 |
| Task-related factors | Problem difficulty, benchmark source, problem domain, vulnerability type, task recency relative to training data, application domain, program size | 11 |
| AI model-related factors | Model/tool type, model version, AI engine, human vs AI authorship, RAG-based mitigation setting | 10 |
| Programming-language factors | Python, Java, JavaScript, C, C++, cross-language comparisons | 8 |
| Interaction factors | Multi-round prompting, user prompting behavior, checking multiple suggestions, human repair, Copilot Chat fixing style, RAG-assisted generation | 7 |

## 5.4 RQ2: Dependent Variables (Quality Outcomes)

RQ2 aimed at identifying and synthesizing the dependent variables, i.e., the software quality outcomes and metrics, that the primary studies used to evaluate AI-generated code. In contrast to RQ1, which focused on influencing factors, this section addressed the quality dimensions that researchers measured when assessing the outputs of AI-assisted code generation tools.

It is important to note that this SLR focused only on quality-related outcome measures. Although some of the reviewed studies also touched on issues such as development time, repair effort, or productivity, those aspects were not the focus of this review and are left for a future SLR dedicated to effort- and productivity-related outcomes.

Across the 24 primary studies, software quality was not treated as a single, uniform concept. Instead, the studies operationalized quality as a multi-dimensional construct, covering aspects such as: (1) functional correctness and validity, (2) security-related quality, (3) maintainability and code smells, (4) complexity and understandability, (5) reliability, robustness, and consistency, and in a smaller number of studies, (6) efficiency or performance-related quality. We discuss each of these categories next.

### 5.4.1 Functional Correctness and Validity

The most frequently studied dependent variable was functional correctness, i.e., whether the generated code satisfied the intended functionality and produced correct outputs under the given inputs and test cases. Several studies operationalized this through benchmark pass rates, execution-based evaluation, correctness ratios, or acceptance rates on online judging platforms.

For example, P3 studied the correctness of GitHub Copilot's code suggestions by executing the generated solutions on LeetCode test cases. P4 also studied correctness at large scale, measuring acceptance rates of Copilot-generated code across 2,033 LeetCode problems and four programming languages. P5 included functional performance through Pass@1 and unit-test success while simultaneously studying broader quality issues. P6 assessed both validity and correctness of GitHub Copilot's outputs on HumanEval, distinguishing between valid code, fully correct code, partially correct code, and incorrect code. P8 studied code correctness in a controlled comparison of ChatGPT and Copilot. P13 examined solution correctness as the probability of finding a correct answer among multiple Copilot suggestions. P14 used code correctness as one of the core dependent variables when comparing ChatGPT, Copilot, and CodeWhisperer. P15 studied correctness when comparing Copilot-generated solutions with human-written solutions. P19 evaluated whether ChatGPT-generated code was functionally correct across algorithmic tasks in multiple languages. P22 also assessed functional correctness using execution-based evaluation on LeetCode tasks. Finally, P24 evaluated correctness of code generated by different AI engines using human-designed test suites.

A related measure in this category was validity, which referred to whether the generated code was syntactically correct, compilable, or executable. P6 explicitly distinguished validity from correctness, and P14 also included validity as a separate dependent variable. In P22, validity-related failures appeared in the form of compilation errors and runtime failures, showing that some studies treated validity as a prerequisite to deeper quality evaluation.

Overall, the literature showed that functional correctness was the most established and widely used quality outcome measure, often serving as the first and most visible indicator of AI-generated code quality.

### 5.4.2 Security-Related Quality

A second major category of dependent variables concerned security-related quality, including vulnerabilities, insecure coding practices, security weaknesses, and security smells. These measures were especially prominent in studies evaluating code generation in security-sensitive contexts.

For example, P2 studied the prevalence of security vulnerabilities in ChatGPT-aided Python code. P5 measured vulnerabilities as part of a broader SonarQube-based assessment of bugs, code smells, and security issues across five LLMs. P9 evaluated security as one of the quality dimensions in a multi-tool comparative study using SonarQube and Jenkins. P10 studied security weaknesses in Python code generated under different prompt optimization strategies. P11 examined security as one of the three core quality dimensions, together with maintainability and reliability, in ChatGPT-generated code from DevGPT. P12 treated security of produced code as the main dependent variable in a user study investigating whether users wrote more insecure code with AI assistants. P14 explicitly included code security as one of the quality metrics when comparing Copilot, CodeWhisperer, and ChatGPT. P16 studied security vulnerabilities using Common Weakness Enumeration (CWE) categories in a large-scale comparison of human-written and AI-generated code. P17 measured whether Copilot reintroduced vulnerable code or generated the corresponding fix in security-related scenarios. P19 evaluated security vulnerabilities in both algorithmic and CWE-based code generation scenarios with ChatGPT. P21 studied both security smells and security-related static-analysis issues in ChatGPT-generated code used by developers. P23 measured the presence and rate of security weaknesses, categorized them using CWE types, and also evaluated fix success when Copilot Chat was used to repair them.

Taken together, these studies showed that security-related quality was one of the most intensively studied outcome areas after correctness. They also showed that the literature increasingly treated security not as a peripheral concern, but as a core dimension of code quality in AI-assisted programming.

#### 5.4.3 Maintainability, Code Smells, and Technical Debt

Another prominent category of dependent variables concerned maintainability and internal code quality, often operationalized through code smells, technical debt, coding-standard violations, or broader maintainability indicators.

P5 studied code smells and issue density through static analysis, explicitly linking them to the broader quality profile of LLM-generated Java code. P9 evaluated maintainability using SonarQube-based quality indicators in its cross-tool comparative study. P11 examined maintainability as one of the three main quality dimensions associated with prompt patterns. P14 explicitly included maintainability and technical debt in its evaluation of Copilot, CodeWhisperer, and ChatGPT. P16 observed maintainability-related issues in both human-written and AI-generated code, especially in relation to structural and defect characteristics. P21 studied code smells in ChatGPT-generated code and also analyzed how much such code was modified before being merged by developers. P22 examined maintainability-related issues through code style and static-analysis findings during refinement of ChatGPT-generated code. P24 also considered software quality metrics and adherence to coding standards, which implicitly reflected maintainability concerns.

In addition, P8 treated "code quality" as a dependent variable in its controlled comparison of ChatGPT and Copilot, and its empirical setup relied on quality rule violations as an operational indicator. Similarly, P7 discussed readability and usability, which, while not always formalized through static metrics, were closely related to maintainability concerns.

In general, these studies showed that maintainability was often assessed through static-analysis proxies rather than direct long-term maintenance studies. This means that the literature typically inferred maintainability from internal code properties such as smells, debt, and standards conformance, rather than from downstream maintenance effort.

#### 5.4.4 Complexity, Understandability, and Structural Characteristics

A fourth category of dependent variables focused on complexity, understandability, and structural characteristics of generated code. These metrics were used to capture internal structural quality and, in some studies, to approximate how easy the code would be to inspect, comprehend, or evolve.

P1 evaluated seven Halstead complexity metrics, including program length, volume, difficulty, effort, time, and delivered bugs, using them as indicators of code quality and complexity in ChatGPT-generated Python snippets. P3 measured cyclomatic complexity and cognitive complexity to evaluate the understandability of Copilot's suggestions. P5 also included cyclomatic and cognitive complexity, along with lines of code and issue density, as part of a broader structural quality analysis. P16 studied structural complexity when comparing human-written and AI-generated code, finding systematic differences in the structural profiles of the two. P19 assessed both cyclomatic and cognitive complexity across code generated by ChatGPT in different languages and under multi-round fixing scenarios. P24 likewise considered software quality metrics that included code quality and structural aspects.

Some studies also used more qualitative notions related to understandability. For example, P7 discussed readability and usability when comparing GitHub Copilot with genetic programming, while P15 considered diversity and reproducibility of Copilot outputs in relation to their practical value for human programmers.

#### 5.4.5 Reliability, Robustness, and Consistency

A fifth category included dependent variables related to reliability, robustness, and consistency. These outcomes captured whether generated code behaved consistently, handled varying inputs well, avoided brittle behavior, or remained stable across repeated or paraphrased prompting.

P1 explicitly studied consistency across code generated by different versions of ChatGPT and across iterative prompts, making consistency itself a core dependent variable. P11 treated reliability as one of the three main quality dimensions in its empirical assessment of prompt patterns. P14 also included reliability as one of the quality metrics when comparing multiple AI-assisted code generation tools. P20 studied robustness by examining whether semantically equivalent paraphrases of the input description changed Copilot's generated code and whether correctness depended on prompt wording. P19 studied the effects of ChatGPT's non-determinism on quality outcomes, including correctness, complexity, and security. P22 examined runtime errors and wrong outputs in refinement scenarios, which also reflected reliability-related concerns.

Some studies additionally captured reliability indirectly through runtime behavior. For example, P22 distinguished compilation errors, runtime errors, and wrong outputs, while P4 identified 17 categories of error reasons behind incorrect code suggestions. These measures did not always use the term "reliability," but they operationalized closely related concepts.

This category suggested that beyond static correctness on a fixed benchmark, some studies increasingly recognized that AI-generated code quality also depends on whether outputs are stable, robust to variation, and consistently reproducible.

### 5.4.6 Efficiency and Performance-Related Quality

A smaller group of studies assessed efficiency or performance-related quality, such as execution time, space complexity, or broader notions of code efficiency.

P6 explicitly included efficiency as one of the three core quality dimensions in its evaluation of GitHub Copilot's outputs. P15 studied efficiency when comparing Copilot-generated code with human solutions. P22 included performance and efficiency issues among the common quality problems identified in ChatGPT-generated code. In P7, execution time and usability concerns also appeared in the discussion when comparing Copilot with genetic programming.

Although this category appeared less frequently than correctness, security, or maintainability, it showed that some researchers treated efficiency as part of software quality, especially when the generated code solved algorithmic tasks where time and space performance mattered.

### 5.4.7 Summary of Quality Outcome Measures

Table 3 summarizes the taxonomy of dependent variables identified across the 24 primary studies. The results showed that functional correctness and validity were the most frequently used outcome measures, followed by security-related quality and maintainability/internal quality measures. Complexity and structural metrics also appeared regularly, often as proxies for understandability and maintainability. In contrast, reliability/robustness/consistency and especially efficiency/performance-related quality were less frequently studied.

**Table 3-Summary of Quality-Related Dependent Variables (RQ2)**

| Quality Outcome Category | Example Metrics / Operationalization | # of Studies | Example Studies |
|---|---|---|---|
| **Functional Correctness & Validity** | Test-case pass rate, acceptance rate, execution correctness, compilation success, runtime success | 18 | P3, P4, P5, P6, P8, P13, P14, P15, P19, P22, P24 |
| **Security-Related Quality** | Vulnerability count, CWE categories, security weaknesses, insecure patterns, security smells, exploitability | 13 | P2, P5, P9, P10, P11, P12, P14, P16, P17, P19, P21, P23 |
| **Maintainability & Code Smells** | Code smells, technical debt, maintainability index, rule violations, static analysis issues, coding standard violations | 12 | P5, P9, P11, P14, P16, P21, P22, P24 |
| **Complexity & Structural Quality** | Cyclomatic complexity, cognitive complexity, Halstead metrics, LOC, structural patterns, readability proxies | 9 | P1, P3, P5, P16, P19, P24 |
| **Reliability, Robustness & Consistency** | Output consistency, robustness to prompt variation, runtime errors, stability across runs, reproducibility | 7 | P1, P11, P14, P19, P20, P22 |
| **Efficiency & Performance** | Execution time, space complexity, algorithmic efficiency, performance bottlenecks | 4 | P6, P7, P15, P22 |

In general, as we can observe, the literature treated software quality in AI-assisted code generation as a multi-dimensional construct rather than a single measurable property. This finding is important because it suggested that conclusions about the "quality" of AI-generated code depend heavily on which dependent variables are selected and how they are operationalized. A tool that performs well in correctness may still exhibit weaknesses in security, maintainability, or consistency. Therefore, future empirical studies should be explicit about the dimensions of quality they measure and should ideally assess multiple complementary dimensions when evaluating AI-generated code.

**Insight:** A key insight from the synthesis is that AI-assisted development shifts code quality from being primarily determined by individual developer capabilities to a socio-technical outcome shaped by the interaction between human inputs, AI system behavior, and evaluation practices.

## 5.5 Insights Regarding Independent versus Dependent Variables, and Formulations of RQs in the Studies

### 5.5.1 Combinations (Pairs) of Independent Variables versus Dependent Variables

To further synthesize the findings of this review, we analyzed how different categories of independent variables (Section 5.3) were studied in relation to specific quality outcomes (Section 5.4). Rather than considering these dimensions in isolation, this analysis examined the pairwise combinations between influencing factors and quality metrics across the primary studies.

Table 4 presents this mapping, where independent variable categories are linked to quality outcome categories, and the frequency of each combination reflects the number of studies that investigated that relationship.

The results showed that AI model characteristics and prompt-related factors were most frequently studied in relation to functional correctness, indicating that the majority of existing research focused on improving or evaluating correctness through model selection and prompt design. Similarly, security-related quality was often examined in combination with prompt characteristics and AI model behavior, reflecting concerns about vulnerability generation and insecure coding patterns.

In contrast, maintainability and complexity-related outcomes were less consistently studied across different independent variables and were primarily associated with AI model characteristics and programming language factors. Reliability and consistency were examined in a smaller subset of studies, often in relation to prompt variation and model stochasticity.

The mapping shown in Table 4 reveals an uneven research landscape, where certain combinations (e.g., prompts–correctness, models–correctness, models–security) were heavily studied, while others (e.g., human–AI interaction vs. maintainability or efficiency) remained underexplored. These gaps suggest opportunities for future research to investigate less-studied combinations and provide a more comprehensive understanding of AI-assisted code quality.

**Table 4-Matrix showing the frequency of studies considering each combination (pair) of independent versus dependent variables**

| Independent Variables | Dependent Variables | | | | |
|---|---|---|---|---|---|
| | Correctness | Security | Maintainability | Complexity | Reliability |
| Prompts | 6 | 4 | 3 | 2 | 3 |
| AI Model | 8 | 5 | 4 | 3 | 4 |
| Tasks | 7 | 3 | 2 | 3 | 2 |
| Programming Language | 5 | 3 | 2 | 2 | 2 |
| Human–AI Interaction | 4 | 2 | 3 | 1 | 3 |

### 5.5.2 RQs Raised and Addressed in the Studies

We analyzed the research questions (RQs) posed in the primary studies (see Appendix X) to understand how the field frames its investigations into AI-assisted code generation and software quality. This analysis reveals clear differences in the types of questions asked, their level of abstraction, and the nature of insights they aim to produce.

A large portion of studies formulated evaluation-focused RQs, focusing on assessing specific quality attributes of AI-generated code. These questions typically examined correctness, validity, or security using quantitative metrics. For example, P14 examined how correct, secure, reliable, and maintainable the outputs of different AI tools were, while P6 evaluated the validity and correctness of GitHub Copilot's code suggestions. Similarly, P3 and P4 assessed correctness using execution-based evaluation on LeetCode tasks, and P22 investigated correctness alongside compilation errors and runtime failures. These RQs reflected a measurement-driven perspective, where quality was operationalized through observable outcomes such as pass rates, defect counts, or failure types.

Another group of studies posed comparative RQs, aiming to evaluate differences across tools, models, or approaches. For instance, P8 compared ChatGPT and GitHub Copilot in terms of code quality, while P14 conducted a broader comparison across ChatGPT, Copilot, and CodeWhisperer. P7 compared GitHub Copilot with genetic programming techniques, and P16 contrasted human-written and AI-generated code at scale. These studies framed their RQs around relative differences across alternatives, often under controlled experimental conditions.

A third category included factor-oriented (explanatory) RQs, which investigated how specific independent variables influenced quality outcomes. For example, P20 examined how variations in prompt phrasing (e.g., paraphrased problem descriptions) affected correctness, while P10 studied how prompt optimization influenced the security of generated code. P13 analyzed how the number of generated suggestions impacted the likelihood of finding a correct solution, and P11 investigated how different prompt patterns affected maintainability, reliability, and security. These RQs moved beyond direct evaluation to explore cause–effect relationships between inputs and outcomes.

Some studies also formulated interaction-oriented RQs, examining how human–AI interaction shaped code quality. For example, P21 investigated how developers used, modified, and integrated AI-generated code in practice, while P12 examined whether users produced more insecure code when assisted by AI tools. P4, although primarily focused on correctness, also provided insights into how developers interact with multiple generated suggestions when selecting solutions. These studies highlighted that code quality is influenced not only by the AI system itself but also by how developers engage with it during the development process.

In addition, several studies adopted exploratory or phenomenon-driven RQs, particularly when investigating emerging behaviors of AI systems. For example, P1 examined the consistency of ChatGPT-generated code across different prompts and model versions, while P19 explored variability and non-determinism in generated outputs across multiple runs. P17 investigated how AI systems handled vulnerable code, including whether they generated fixes or reintroduced vulnerabilities. These RQs aimed to characterize emerging and not yet fully understood phenomena, rather than evaluate predefined hypotheses.

#### 5.5.2.1 RQ Typology Identified in the Literature

Based on this analysis, the RQs in the primary studies can be categorized into the typology shown in Table 5.

**Table 5- RQ Typology Identified in the Literature**

| RQ Type | Description | Example Studies |
|---|---|---|
| **Evaluation-focused** | Assess quality outcomes such as correctness, security, or maintainability | P3, P4, P6, P14, P22 |
| **Comparative** | Compare different tools, models, or approaches | P7, P8, P14, P16 |
| **Factor-oriented (Explanatory)** | Analyze how specific factors influence quality outcomes | P10, P11, P13, P20 |
| **Interaction-oriented** | Examine how human–AI interaction affects code quality | P12, P21 |
| **Exploratory / Phenomenon-driven** | Investigate emerging behaviors such as hallucinations or inconsistency | P1, P17, P19 |

This categorization shows that the literature spans multiple levels of inquiry, ranging from direct measurement of quality outcomes to comparative evaluation, and increasingly toward explanatory and interaction-focused investigations. The distribution of RQ types also indicates that most studies concentrate on evaluating the quality of AI-generated code, while fewer studies address deeper causal mechanisms or the broader implications of human–AI collaboration in software development.

> **Insight:** The findings suggest that many quality issues in AI-generated code do not arise solely from human or AI errors, but from misalignment between developer intent and AI-generated outputs, highlighting the central role of human–AI interaction in defect introduction.

## 5.6 RQ3: Types of Evaluation Methods

RQ3 examined the empirical evaluation approaches, procedures, tools, and analysis methods used in the primary studies to assess the quality of AI-generated code. Understanding these evaluation methods is important because the validity, reliability, and comparability of findings depend heavily on how quality is measured and assessed. Different evaluation approaches may emphasize different aspects of quality (e.g., correctness vs. maintainability), which in turn shapes the conclusions drawn in the literature.

The analysis of the primary studies revealed that researchers employed a diverse set of evaluation methods, which can be grouped into five main categories: (1) benchmark-based evaluation, (2) testing-based evaluation, (3) static analysis, (4) human-centered studies, and (5) statistical analysis methods.

### 5.6.1 Benchmark-Based Evaluation

A large number of studies relied on benchmark datasets and coding platforms to evaluate AI-generated code in a standardized and reproducible manner. These benchmarks typically consisted of programming problems with predefined test cases, enabling automated correctness evaluation.

For example, P4 evaluated GitHub Copilot using 2,033 LeetCode[1] problems across multiple programming languages, while P3 also used LeetCode tasks to assess correctness and complexity. P6 and P13 employed the HumanEval[2] benchmark to evaluate functional correctness and solution diversity. P19 and P22 similarly used benchmark-style algorithmic tasks to assess correctness, complexity, and security properties.

In addition to standardized benchmarks, some studies used mined datasets from real-world sources. For instance, P11 analyzed code and prompts from the DevGPT dataset [72] to study quality dimensions such as maintainability, reliability, and security. These approaches enabled large-scale evaluation under realistic conditions.

### 5.6.2 Testing-Based Evaluation

Another widely used approach was testing-based evaluation, where generated code was executed against test cases to assess correctness and runtime behavior.

Several studies used unit testing and execution-based validation. For example, P6 evaluated Copilot-generated code using execution testing and classified outputs into correct, partially correct, and incorrect. P22 assessed generated code using compilation, runtime, and output-based testing, identifying issues such as runtime errors and incorrect outputs. P24 used human-designed test suites to evaluate correctness across different AI-generated solutions.

Some studies also employed framework-based testing, such as JUnit or structured test harnesses, particularly when evaluating object-oriented or multi-class systems (e.g., P5 and P9). These approaches ensured systematic and repeatable evaluation of functional behavior.

### 5.6.3 Static Analysis

A significant portion of studies used static analysis tools to evaluate internal code quality attributes such as maintainability, complexity, and security vulnerabilities.

For example, P5 and P9 used the SonarQube[3] commercial tool to analyze bugs, code smells, and security issues in AI-generated code. P1 and P19 used Radon to compute complexity metrics such as cyclomatic complexity and Halstead metrics. P8 and P21 applied rule-based static analysis to identify quality violations and code smells. P24 also incorporated static analysis tools to assess adherence to coding standards and structural quality.

Static analysis enabled the evaluation of non-functional quality attributes that are not easily captured through execution-based testing, particularly maintainability and internal code quality.

### 5.6.4 Human-Centered Studies

Some studies adopted human-centered evaluation approaches, focusing on how developers interact with AI-generated code and how this interaction affects quality.

For example, P12 conducted a controlled user study to examine whether developers produced more insecure code when assisted by AI tools. P21 analyzed developer activity and repository data to study how AI-generated code was modified and integrated in practice. P4 included aspects of developer interaction when analyzing how multiple suggestions were explored and selected.

These approaches provided insights into real-world usage and human–AI collaboration, which cannot be captured through automated evaluation alone.

---

[1] LeetCode (leetcode.com) is an online platform for coding interview preparation. The platform provides coding and algorithmic problems intended for users to practice coding.

[2] github.com/openai/human-eval

[3] sonarsource.com/products/sonarqube/

#### 5.6.5 Statistical Analysis Methods

Across the studies, various statistical analysis methods were used to interpret evaluation results and support empirical conclusions.

Most studies employed descriptive statistics, such as means, medians, and distributions, to summarize correctness rates, vulnerability counts, and complexity measures (e.g., P3, P4, P6, P19). Several studies also applied inferential statistical tests, such as the Mann–Whitney U test, significance testing, and hypothesis testing, to compare different tools or conditions (e.g., P8, P14, P16).

These statistical techniques were essential for determining whether observed differences were meaningful and for supporting claims about the relative quality of AI-generated code.

#### 5.6.6 Summary of Evaluation Methods

The analysis shows that the literature relies on a combination of automated, analytical, and human-centered evaluation approaches (Table 6). Benchmark-based and testing-based methods dominate the evaluation of correctness, while static analysis is commonly used for maintainability and security. Human-centered studies remain less frequent but provide important insights into real-world usage and interaction dynamics.

The diversity of evaluation methods highlights that assessing AI-generated code requires multiple complementary approaches, as no single method captures all dimensions of software quality. It also suggests the need for more standardized and integrated evaluation frameworks to improve comparability across studies.

**Table 6-Summary of Evaluation Methods in the Primary Studies**

| Evaluation Category | Tools / Techniques | Example Studies |
|---|---|---|
| Benchmark-Based Evaluation | LeetCode, HumanEval, DevGPT, coding benchmarks | P3, P4, P6, P11, P13, P19, P22 |
| Testing-Based Evaluation | Unit testing, execution testing, JUnit, test suites | P5, P6, P9, P22, P24 |
| Static Analysis | SonarQube, Radon, CheckStyle, code smell detection | P1, P5, P8, P9, P19, P21, P24 |
| Human-Centered Studies | User studies, developer activity analysis, repository mining | P4, P12, P21 |
| Statistical Analysis | Descriptive statistics, Mann–Whitney U test, significance testing | P3, P4, P8, P14, P16, P19 |

## 5.7 RQ4: Datasets and Programming Tasks

For RQ4, we examined the datasets, benchmarks, and programming tasks used in the primary studies to evaluate AI-generated code. The choice of datasets and tasks plays a central role in shaping evaluation outcomes, as it determines the type of problems addressed, the realism of the evaluation setting, and the dimensions of quality being assessed.

The analysis showed that the primary studies relied on a combination of synthetic (benchmark-based) datasets, real-world datasets, and security-oriented datasets, alongside a range of programming task types. In this paper, we distinguish between synthetic datasets, which consist of curated or artificially constructed programming problems designed for controlled evaluation, and real-world datasets, which are derived from actual developer activity or software repositories.

#### 5.7.1 Synthetic (Benchmark-Based) Datasets

A large proportion of studies used synthetic datasets, i.e., curated collections of programming problems designed specifically for evaluating code generation systems. These datasets are typically constructed with predefined inputs, expected outputs, and test cases, enabling controlled and reproducible evaluation.

For example, LeetCode datasets were widely used to evaluate correctness in algorithmic problem-solving scenarios. P3 and P4 evaluated GitHub Copilot using large sets of LeetCode problems, while P22 used LeetCode-style tasks to analyze correctness, compilation errors, and runtime failures. These datasets consist of well-defined algorithmic challenges and are not derived from real software projects, which is why they are considered synthetic.

Similarly, HumanEval was used in P6 and P13 to evaluate correctness and solution diversity. HumanEval is a curated benchmark of Python programming problems designed specifically for assessing code generation models. P19 also used algorithmic tasks similar to HumanEval-style problems when evaluating ChatGPT across multiple languages.

In addition, broader benchmark suites such as CodeXGLUE[1] were used in studies such as P5 and P14 to evaluate code generation across multiple tasks. These benchmarks aggregate curated datasets for different code intelligence tasks and provide standardized evaluation settings.

These synthetic datasets primarily supported controlled experimentation, enabling direct comparison across models and configurations. However, they typically focused on isolated algorithmic tasks, which may not fully reflect real-world software development complexity.

#### 5.7.2 Real-World Datasets and Repository-Based Data

In contrast, several studies used real-world datasets, derived from actual developer activity or software repositories. These datasets captured realistic coding scenarios, including incomplete specifications, evolving codebases, and human–AI interaction.

For example, P11 used the DevGPT dataset, which contains real prompts and generated code collected from developers using AI tools. This dataset enabled the study of maintainability, reliability, and security in realistic usage contexts. Similarly, P21 analyzed GitHub repositories to examine how developers modified and integrated AI-generated code in practice. P16 also relied on large-scale datasets of human-written and AI-generated code to compare their structural and quality characteristics.

Unlike synthetic datasets, these datasets are not artificially constructed for evaluation purposes, but instead reflect actual development practices, making them more representative of real-world usage. At the same time, they introduce greater variability and less controlled conditions compared to benchmark datasets.

#### 5.7.3 Security-Oriented Datasets

A subset of studies used security-focused datasets, particularly those based on known vulnerability patterns and standardized taxonomies.

For example, P17 and P23 used CWE (Common Weakness Enumeration) scenarios to evaluate whether AI systems generated vulnerable code or were able to produce secure fixes. P19 also incorporated CWE-based scenarios alongside algorithmic tasks to assess security vulnerabilities in generated code. P2 and P12 similarly evaluated security-related outcomes, including insecure coding practices and vulnerability introduction in AI-assisted code.

These datasets enabled targeted evaluation of security-related quality dimensions, which are not typically covered by general-purpose coding benchmarks.

#### 5.7.4 Programming Task Types

Beyond datasets, the primary studies differed in the types of programming tasks used for evaluation. These tasks can be grouped into several categories:

- Algorithmic problem-solving tasks: Widely used in P3, P4, P6, P13, P19, and P22, where AI systems generated solutions to well-defined problems with clear correctness criteria. These tasks were typically associated with synthetic datasets such as LeetCode and HumanEval.
- API usage and code synthesis tasks: Evaluated in studies such as P5 and P9, where AI systems generated code involving library usage, structured programs, or multi-step implementations.
- Bug fixing and vulnerability repair tasks: Studied in P17 and P23, where AI systems were evaluated on their ability to fix vulnerabilities or produce secure code in CWE-based scenarios.
- Code completion and suggestion tasks: Investigated in P3, P4, and P13, where AI systems generated code based on partial inputs or prompts, reflecting common usage of tools such as GitHub Copilot.
- Real-world development tasks: Observed in P11 and P21, where tasks emerged from actual developer workflows, including feature implementation, modification, and integration of AI-generated code.

These task types ranged from well-defined and constrained problems to open-ended and context-rich development activities, influencing both the evaluation methods used and the quality outcomes observed.

---

[1] github.com/microsoft/codexglue

Table X.

#### 5.7.5 Summary of Datasets and Tasks

As summarized in Table 7, the analysis of RQ4 shows a clear distinction between synthetic benchmark datasets, which support controlled and reproducible evaluation, and real-world datasets, which capture practical development scenarios and human–AI interaction. Synthetic datasets dominate the literature, particularly for correctness evaluation (e.g., P3, P4, P6, P13), while real-world datasets (e.g., P11, P16, P21) remain less frequently used but provide more realistic insights.

Security-oriented datasets (e.g., P17, P19, P23) form a specialized category focused on vulnerability analysis. Across all categories, algorithmic tasks are the most common, whereas real-world development tasks are comparatively underexplored. This distribution indicates that current evaluations are often conducted under controlled conditions, with fewer studies addressing the complexity of real-world software engineering contexts.

**Table 7-Summary of Datasets and Programming Tasks**

| Dataset / Source | Type | Task Type | Example Studies |
|---|---|---|---|
| LeetCode | Synthetic (benchmark) | Algorithmic problems, code completion | P3, P4, P22 |
| HumanEval | Synthetic (benchmark) | Algorithmic tasks, correctness evaluation | P6, P13 |
| CodeXGLUE / Benchmarks | Synthetic (benchmark) | Multi-task evaluation | P5, P14 |
| DevGPT Dataset | Real-world | Real developer prompts and code | P11 |
| GitHub Repositories | Real-world | Developer activity, real coding tasks | P16, P21 |
| CWE Scenarios | Security-focused | Vulnerability detection and repair | P17, P19, P23 |
| Custom / Study-specific Tasks | Mixed | API usage, synthesis, testing | P8, P9, P24 |

## 5.8 Synthesis of Empirical Evidence

### 5.8.1 RQ5: Reported Empirical Evidence (Impact on Code Quality)

RQ5 examined the empirical evidence reported in the primary studies regarding the impact of AI-assisted code generation on software code quality. Following established practices in evidence synthesis, the findings were aggregated and interpreted across studies, rather than reported in isolation, to identify recurring patterns, conditions, and limitations. Given the diversity of evaluation methods (Section 5.6), datasets and tasks (Section 5.7), and quality metrics (Section 5.3), the evidence is synthesized into three categories: positive, mixed, and negative findings. These categories represent patterns of evidence, not absolute or directly comparable outcomes.

#### 5.8.1.1 Positive Findings

Several studies reported favorable outcomes under specific and controlled conditions, particularly for well-defined programming tasks.

For example, P3, P4, and P6 showed that AI-generated code was often able to produce test-passing solutions in benchmark settings, especially for algorithmic problems drawn from datasets such as LeetCode and HumanEval. However, these outcomes depended on the adequacy of test suites and the nature of the tasks, as test-passing behavior does not necessarily imply full semantic correctness or robustness.

P13 further demonstrated that considering multiple generated candidates increased the likelihood of obtaining a correct solution, indicating that quality can be improved through selection strategies rather than relying on single outputs. Comparative studies such as P8 and P14 reported that different AI tools were capable of generating functionally adequate code across a range of tasks, suggesting a baseline level of competence across modern systems.

In addition, studies such as P21 indicated that AI-generated code could be useful in practical development workflows, supporting developers in producing working code more efficiently, although such benefits were typically observed alongside the need for human validation and modification.

Taken together, these findings suggest that AI-generated code can achieve acceptable functional quality in constrained and well-specified scenarios, particularly when supported by appropriate prompts, multiple candidate generation, and human oversight.

#### 5.8.1.2 Mixed Findings

A substantial portion of the literature reported context-dependent and variable outcomes, indicating that the quality of AI-generated code is highly sensitive to multiple interacting factors.

For instance, P4 and P22 showed that correctness varied significantly with problem difficulty, with relatively strong results on simpler tasks but substantial degradation on more complex or less common problems. Similarly, P6 reported a mix of fully correct, partially correct, and incorrect outputs, highlighting variability even within the same evaluation setting.

Several studies identified limitations in internal code quality, even when functional correctness was achieved. For example, P1 and P19 observed that AI-generated code could exhibit higher complexity, redundancy, or suboptimal structure, affecting maintainability. P11 further showed that maintainability, reliability, and security outcomes varied depending on prompt formulation and context.

Another recurring finding was inconsistency across repeated generations. P1 and P19 reported that identical prompts could produce different outputs, raising concerns about reproducibility and reliability. P13 also highlighted that while generating multiple candidates improves the chance of correctness, it introduces greater variability, requiring additional effort for selection and validation.

These findings indicate that the impact of AI-generated code on quality is non-uniform and highly dependent on task characteristics, prompt design, and evaluation conditions, rather than being consistently positive or negative.

#### 5.8.1.3 Negative Findings

Several studies reported systematic limitations and risks, particularly in relation to correctness in complex scenarios, security vulnerabilities, and reliability.

One commonly reported issue was the generation of hallucinated or non-existent APIs, as observed in P1 and P19. In these cases, the AI produced syntactically plausible but invalid code that referenced functions or libraries that do not exist, leading to subtle correctness issues that may not be detected by basic tests.

Security concerns were also prominent. P12 showed that developers using AI assistance were more likely to produce insecure code, while P17 and P23 demonstrated that AI-generated code could contain known vulnerabilities or fail to correctly address them in CWE-based scenarios. P2 similarly identified insecure coding patterns in AI-generated outputs.

In addition, studies such as P6 and P22 reported compilation errors, runtime failures, and incorrect outputs, particularly for more complex tasks. P1 and P19 further highlighted non-deterministic and inconsistent behavior, where repeated executions produced different results, undermining reliability.

These findings suggest that AI-generated code can introduce hidden defects, security risks, and reliability issues, especially when used without rigorous validation and human oversight.

#### 5.8.1.4 Synthesis of Evidence

Table 8 presents a pattern-based synthesis of the empirical evidence reported across the primary studies, organizing findings along key code quality dimensions rather than listing individual results. The table highlights that the impact of AI-generated code is highly conditional, with positive outcomes—such as test-passing functionality and practical usefulness—primarily observed in well-defined tasks and controlled evaluation settings. At the same time, a substantial body of evidence points to variability and trade-offs, where functional correctness does not necessarily translate into robust, maintainable, or consistent code. The synthesis further reveals recurring risks, including security vulnerabilities, hallucinated APIs, and non-deterministic outputs, particularly in complex or poorly specified scenarios. By explicitly linking each pattern to its typical context and supporting studies, the table provides a structured view of where AI-generated code performs well, where it degrades, and under what conditions these outcomes emerge, offering a more nuanced understanding than study-level summaries.

**Table 8-Synthesized Empirical Evidence on AI-Generated Code Quality**

| Evidence Category | Quality Aspect | Observed Pattern Across Studies | Typical Conditions / Context | Representative Studies |
|---|---|---|---|---|
| **Positive** | Functional correctness | Test-passing solutions achievable under controlled conditions | Simple / well-defined tasks (e.g., algorithmic problems, strong test suites) | P3, P4, P6, P13 |
| | Usability / practical utility | Generated code is usable with minor modifications | Developer-in-the-loop scenarios | P7, P21 |
| **Mixed** | Correctness robustness | Correctness varies significantly across tasks and inputs | Performance degrades with task complexity or ambiguity | P4, P22, P20 |
| | Internal code quality | Code may be functionally correct but structurally suboptimal | High complexity, redundancy, maintainability issues | P1, P5, P11 |
| | Consistency / determinism | Outputs vary across repeated runs for the same prompt | Non-deterministic generation behavior | P1, P19 |
| | Multi-solution generation | Multiple outputs improve success probability but increase variability | Requires human selection/filtering | P13 |
| **Negative** | Security | AI-generated code may include vulnerabilities or insecure patterns | Security-sensitive tasks, weak prompts | P2, P12, P17, P23 |
| | Hallucinations | Non-existent APIs or incorrect library usage generated | Knowledge gaps, ambiguous prompts | P1, P19 |
| | Reliability | Compilation errors, runtime failures, incomplete solutions | Complex or less common tasks | P6, P22 |
| | Overconfidence risk | Code appears plausible but is incorrect or unsafe | Lack of validation or human oversight | P12, P19 |

### 5.8.2 RQ6: Synthesis of Reported AI–Human Collaboration Best Practices to Achieve High-Quality Code

RQ6 examined the AI–human collaboration practices reported in the primary studies that contribute to improving the quality of AI-generated code. Rather than treating AI as a fully autonomous code generator, the literature consistently emphasizes collaborative processes, where human input, oversight, and iteration play a critical role in achieving reliable outcomes.

The extracted evidence reveals several recurring categories of best practices, including prompt engineering, iterative interaction, human validation, and hybrid development workflows. These practices reflect a shift from one-shot code generation toward interactive and guided use of AI systems.

#### 5.8.2.1 Prompt Engineering and Input Quality

A central finding across multiple studies is the importance of well-structured and precise prompts. Studies such as P10 and P20 show that detailed, context-rich prompts—including explicit requirements, constraints, and examples—lead to more accurate and secure outputs.

Conversely, vague or underspecified prompts were associated with incorrect, incomplete, or insecure code (e.g., P2, P11). This highlights that prompt formulation acts as a critical control mechanism, directly influencing the quality of generated code.

#### 5.8.2.2 Iterative and Multi-Round Interaction

Several studies emphasize the effectiveness of iterative interaction, where developers refine prompts and outputs over multiple rounds. P13 demonstrated that generating and evaluating multiple candidate solutions increases the likelihood of obtaining correct results, while P20 showed that incremental prompt refinement improves both correctness and robustness.

This iterative process allows developers to progressively guide the AI, correct misunderstandings, and explore alternative implementations, mitigating limitations of single-shot generation.

#### 5.8.2.3 Human Review and Validation

A consistent recommendation across the literature is the necessity of human oversight. Studies such as P12, P17, and P23 highlight that AI-generated code may contain security vulnerabilities, hallucinations, or subtle defects, which are not reliably detected without human inspection.

As a result, developers are expected to perform code review, testing, and validation, particularly for security-sensitive or production-critical code. Human involvement is therefore essential not only for correctness but also for ensuring robustness and trustworthiness.

#### 5.8.2.4 Hybrid AI–Human Collaboration

The literature converges on the effectiveness of hybrid workflows, where AI and humans contribute complementary strengths. In this model, AI systems are typically used to generate initial solutions, boilerplate code, or alternative implementations, while humans are responsible for refinement, optimization, and validation.

Studies such as P7 and P21 indicate that this division of labor improves development efficiency while maintaining acceptable quality levels. However, the effectiveness of such workflows depends on active human engagement, rather than passive acceptance of AI outputs.

#### 5.8.2.5 Awareness of Limitations and Risk Mitigation

Another important practice is maintaining awareness of AI limitations, including non-determinism, hallucinations, and security risks. Studies such as P1 and P19 suggest that developers should adopt a skeptical and verification-oriented mindset, treating AI-generated code as a draft rather than a final solution.

This includes practices such as: validating API usage, testing edge cases, and verifying assumptions embedded in generated code.

#### 5.8.2.6 Synthesis of Findings

The synthesized evidence (Table 9) indicates that high-quality AI-generated code is not achieved through AI automation alone, but through structured and active collaboration between developers and AI systems. Practices such as prompt engineering and iterative interaction directly influence the inputs and outputs of the AI, while human review and hybrid workflows ensure that generated code meets broader quality requirements.

These findings reinforce that, as discussed widely in the software industry, AI tools function most effectively as assistive partners rather than autonomous developers, and that achieving high-quality outcomes depends on how developers interact with and govern these tools, rather than the tools alone.

**Table 9-Synthesized AI–Human Collaboration Practices for High-Quality Code**

| Collaboration Practice | Purpose | Observed Benefit | Representative Studies |
|---|---|---|---|
| Prompt engineering | Provide clear and precise task specification | Improves correctness and reduces ambiguity | P10, P20, P11 |
| Iterative interaction | Refine outputs through multi-round prompting | Increases correctness and robustness | P13, P20 |
| Multi-candidate generation | Explore alternative solutions | Improves likelihood of correct solutions | P13 |
| Human review and validation | Detect defects, vulnerabilities, hallucinations | Ensures reliability and security | P12, P17, P23 |
| Hybrid workflows | Combine AI generation with human refinement | Balances efficiency and quality | P7, P21 |
| Risk-aware usage | Recognize AI limitations and verify outputs | Reduces hidden defects and misuse | P1, P19 |
| Testing and verification | Validate functional correctness and edge cases | Improves robustness and dependability | P6, P22 |

## 6 Discussions

### 6.1 Implications of the Findings

The findings of this study have several important implications for both research and practice in AI-assisted software development.

First, the results highlight that software code quality in AI-assisted development is a multi-dimensional outcome shaped by the interplay of human factors, AI system characteristics, and human–AI interaction dynamics. This implies that improving code quality cannot rely solely on advances in AI models; equal attention must be given to prompt design, developer expertise, and effective collaboration workflows. For practitioners, this underscores the importance of developing skills in prompt engineering and critical evaluation of AI-generated outputs.

Second, the synthesis of independent variables (RQ1) reveals that prompt quality and task specification play a central role in influencing the quality of generated code. Ambiguous or incomplete prompts can propagate errors through the AI system, while well-structured inputs can significantly improve outcomes. This suggests that organizations adopting AI coding tools should establish guidelines and best practices for prompt formulation and usage.

Third, the analysis of dependent variables (RQ2) indicates that AI-generated code exhibits variability across multiple quality dimensions, including correctness, security, maintainability, and complexity. While some studies report comparable or even improved performance relative to human-written code, others identify increased risks, particularly in security and robustness. This inconsistency implies that AI-generated code should not be assumed to be reliable by default and must be subjected to rigorous validation and testing processes.

Fourth, the findings emphasize the critical role of human oversight and validation. Across studies, effective review, testing, and iterative refinement are essential to mitigate defects introduced by both human and AI sources. This reinforces the view of AI as a supporting tool rather than a replacement for human developers, requiring continuous human-in-the-loop control.

Fifth, the diversity of evaluation methods, datasets, and tasks observed in the literature suggests a lack of standardization in assessing AI-generated code quality. This limits comparability across studies and indicates a need for more standardized benchmarks, evaluation frameworks, and reporting practices in future research.

Finally, the identified patterns of defects and quality issues suggest that AI-assisted development introduces new categories of risks, particularly related to misalignment between human intent and AI outputs, as well as systematic errors arising from model limitations. Addressing these risks requires not only technical improvements in AI systems but also process-level adaptations, including better integration of AI tools into software engineering workflows and clearer accountability mechanisms.

In summary, these implications reinforce that AI-assisted code generation is not merely a productivity tool but a socio-technical paradigm shift, requiring coordinated advances in tools, practices, and evaluation methods to ensure high-quality software outcomes.

> **Insight:** The results indicate that improving AI-generated code quality requires not only better models, but also better interaction design, prompting strategies, and validation workflows.

## 6.2 Limitations and Threats to Validity

Given the empirical nature of this research, the findings should be interpreted in light of several limitations and potential threats to validity. Like any SLR study, the work reported in this paper is subject to constraints related to context, methodology, and data sources. In the following, we discuss these issues using four commonly adopted categories of validity threats [73]—internal validity, construct validity, conclusion validity, and external validity—and outline steps taken to mitigate them where possible.

**Internal validity.** Threats to internal validity relate to potential biases in the study selection, data extraction, and synthesis processes. The use of AI assistance may introduce risks such as misclassification or incomplete interpretation of primary studies. To mitigate this, all AI-generated outputs were carefully inspected and validated by human researchers, and inclusion/exclusion decisions were made manually. Additionally, full-text verification was performed to ensure accurate interpretation of the selected studies.

**Construct validity.** Construct validity concerns whether the extracted variables and measures accurately represent the concepts of interest, particularly software quality. Given the diversity of definitions and metrics used across primary studies, there is a risk of inconsistency in how quality attributes are interpreted. To address this, we systematically categorized dependent variables into well-defined quality dimensions (e.g., correctness, security, maintainability, complexity), aligning them with commonly accepted software quality perspectives.

**Conclusion validity.** Threats to conclusion validity arise from the heterogeneity of the primary studies, including variations in evaluation methods, datasets, tasks, and reported metrics. This diversity limits the ability to perform statistical aggregation and may affect the strength of inferred relationships. To mitigate this, we adopted a qualitative, pattern-based synthesis approach, focusing on identifying consistent trends across studies rather than relying on quantitative aggregation.

**External validity.** External validity refers to the generalizability of the findings. The selected studies cover a range of AI tools, tasks, and evaluation settings; however, they may not fully represent all real-world development contexts or emerging AI systems. In particular, the rapid evolution of AI models and tools may limit the longevity of some findings. To mitigate this, we included recent studies and multiple data sources, aiming to provide a representative and up-to-date overview of the field.

Overall, while these limitations should be considered when interpreting the results, the systematic methodology, combined with human oversight and validation, helps ensure the reliability and relevance of the findings.

## 7 Conclusions and Future Work

This paper presented a systematic literature review (SLR) on the impact of AI-assisted code generation on software code quality. By synthesizing evidence from 24 empirical studies, the review provided a structured analysis of both independent variables (influencing factors) and dependent variables (quality outcomes), as well as the relationships between them.

The findings show that code quality in AI-assisted software development is shaped by a combination of human factors, AI system characteristics, and human–AI interaction dynamics. While AI tools can improve productivity and, in some cases, code quality, the results across studies are mixed and context-dependent. In particular, variability in outcomes across dimensions such as correctness, security, maintainability, and complexity highlights that AI-generated code cannot be assumed to be reliable without proper validation. The study also identified that prompt design, task specification, and human oversight play critical roles in determining the quality of generated code.

Furthermore, the review revealed a lack of standardization in evaluation methods, datasets, and reporting practices, which limits comparability across studies and complicates evidence synthesis. This fragmentation indicates a need for more consistent and rigorous evaluation frameworks in future research.

Based on these findings, several directions for future work emerge.

- First, while several common benchmarks and evaluation practices are beginning to emerge, the field still lacks widely adopted and consistently applied standards. Further efforts are needed to consolidate these emerging approaches into more standardized and comparable evaluation frameworks for assessing AI-generated code quality across studies.
- Second, future research should explore systematic approaches to prompt engineering and human–AI collaboration, aiming to establish best practices that improve reliability and reduce defect introduction.
- Third, more empirical studies are needed to investigate long-term and real-world usage scenarios, including industrial settings and complex software systems.
- Finally, advancing techniques for detecting and mitigating AI-induced defects, particularly in areas such as security and robustness, remains an important research challenge.

This study highlighted that AI-assisted code generation represents a paradigm shift in software engineering, requiring not only improvements in AI models but also careful integration into development processes to ensure high-quality and trustworthy software systems.

## Appendix A- Additional details (log) of the AI-assisted MLR with Human Oversight and Inspection

### A.1- Interaction with and prompting to the GenAI during the Planning of the SLR

**Prompts given to the AI to start the first initial version of the SLR**

Okay, I think now you can get ready to start the SLR execution, right? So I want you to search, take your time, you can spend 20 or 30 minutes, as much time as you want. I'm gonna leave it. And remember that in the paper, I'm gonna include a meta section about how I managed your work to do AI-assisted SLR, okay? You have the five RQs, you know the scope, you know the inclusion-exclusion criteria. Tell them to me so that I verify what you are gonna do, and then you can start your deep research agentic mode to search for academic papers in this area and extract findings for all those five research questions.

Thought ›

Yes — I am ready to execute **SLR 1**.

Before I start the search, here is the **exact protocol** I will follow, so you can verify that it matches your intention.

#### Final RQs for SLR 1

**RQ1 — Independent variables studied**
What independent variables have been investigated in empirical studies evaluating the impact of AI-based code generation on software code quality?

**RQ2 — Dependent variables (code quality metrics)**
What dependent variables and software code quality metrics are used to evaluate AI-generated code?

**RQ3 — Evaluation methods**
What empirical evaluation methods are used to assess the quality of AI-generated code?

**RQ4 — Datasets and programming tasks used**
What datasets, benchmarks, and programming tasks are used to evaluate AI-generated code?

**RQ5 — Reported evidence (impact on code quality)**
What evidence do empirical studies report regarding the impact of AI-based code generation on software code quality?

## A.2- Interaction with and prompting to the GenAI during the <u>Paper Search Phase</u>

### Searching for the related sources

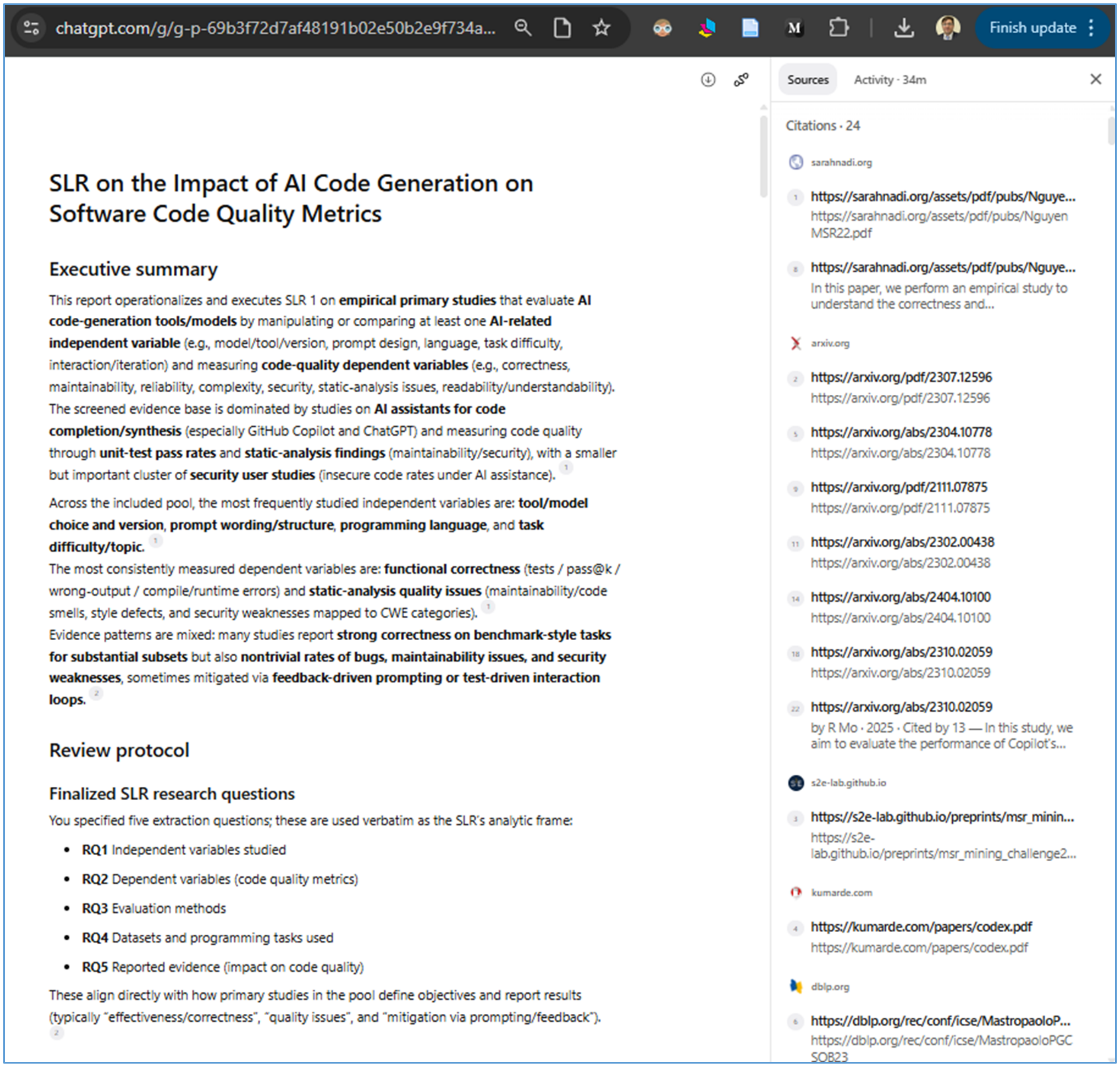

chatgpt.com/g/g-p-69b3f72d7af48191b02e50b2e9f734a...

Finish update

# SLR on the Impact of AI Code Generation on Software Code Quality Metrics

## Executive summary

This report operationalizes and executes SLR 1 on **empirical primary studies** that evaluate **AI code-generation tools/models** by manipulating or comparing at least one **AI-related independent variable** (e.g., model/tool/version, prompt design, language, task difficulty, interaction/iteration) and measuring **code-quality dependent variables** (e.g., correctness, maintainability, reliability, complexity, security, static-analysis issues, readability/understandability). The screened evidence base is dominated by studies on **AI assistants for code completion/synthesis** (especially GitHub Copilot and ChatGPT) and measuring code quality through **unit-test pass rates** and **static-analysis findings** (maintainability/security), with a smaller but important cluster of **security user studies** (insecure code rates under AI assistance). [1]

Across the included pool, the most frequently studied independent variables are: **tool/model choice and version**, **prompt wording/structure**, **programming language**, and **task difficulty/topic**. [1]
The most consistently measured dependent variables are: **functional correctness** (tests / pass@k / wrong-output / compile/runtime errors) and **static-analysis quality issues** (maintainability/code smells, style defects, and security weaknesses mapped to CWE categories). [1]
Evidence patterns are mixed: many studies report **strong correctness on benchmark-style tasks for substantial subsets** but also **nontrivial rates of bugs, maintainability issues, and security weaknesses**, sometimes mitigated via **feedback-driven prompting or test-driven interaction loops**. [2]

## Review protocol

### Finalized SLR research questions

You specified five extraction questions; these are used verbatim as the SLR's analytic frame:

- **RQ1** Independent variables studied
- **RQ2** Dependent variables (code quality metrics)
- **RQ3** Evaluation methods
- **RQ4** Datasets and programming tasks used
- **RQ5** Reported evidence (impact on code quality)

These align directly with how primary studies in the pool define objectives and report results (typically "effectiveness/correctness", "quality issues", and "mitigation via prompting/feedback"). [2]

Sources | Activity · 34m

Citations · 24

sarahnadi.org
1 https://sarahnadi.org/assets/pdf/pubs/Nguye...
https://sarahnadi.org/assets/pdf/pubs/Nguyen MSR22.pdf
8 https://sarahnadi.org/assets/pdf/pubs/Nguye...
In this paper, we perform an empirical study to understand the correctness and...

arxiv.org
2 https://arxiv.org/pdf/2307.12596
https://arxiv.org/pdf/2307.12596
5 https://arxiv.org/abs/2304.10778
https://arxiv.org/abs/2304.10778
9 https://arxiv.org/pdf/2111.07875
https://arxiv.org/pdf/2111.07875
11 https://arxiv.org/abs/2302.00438
https://arxiv.org/abs/2302.00438
14 https://arxiv.org/abs/2404.10100
https://arxiv.org/abs/2404.10100
18 https://arxiv.org/abs/2310.02059
https://arxiv.org/abs/2310.02059
22 https://arxiv.org/abs/2310.02059
by R Mo · 2025 · Cited by 13 — In this study, we aim to evaluate the performance of Copilot's...

s2e-lab.github.io
3 https://s2e-lab.github.io/preprints/msr_minin...
https://s2e-lab.github.io/preprints/msr_mining_challenge2...

kumarde.com
4 https://kumarde.com/papers/codex.pdf
https://kumarde.com/papers/codex.pdf

dblp.org
6 https://dblp.org/rec/conf/icse/MastropaoloP...
https://dblp.org/rec/conf/icse/MastropaoloPGCSOB23

### GenAI was not able to access full-text of a number of papers during the initial search

| Paper title | | RQs (verbatim if present; otherwise | RQ1 Independent variables studied | RQ2 Dependent variables (code- | RQ3 Evaluation methods | RQ4 Dataset / tasks | RQ5 Reported evidence on code |
|---|---|---|---|---|---|---|---|
| Is GitHub's Copilot as Bad as Humans at Introducing Vulnerabilities in Code? | GenAI was not able to access paper's full-text | Objective statement: compare vulnerability introduction tendencies of Copilot vs humans. | **Code source/author (Copilot vs human); language-specific comparisons (study-dependent).** | **Security vulnerabilities (CWE-classed weaknesses) and comparative vulnerability rates.** | Static-analysis/benchmark-based security evaluation (details not extracted here). | Unspecified here (Springer journal article). | Included as archival evidence directly addressing Copilot vs human vulnerability introduction; full extraction required in later phase. [32] |

**Human researchers provided to the GenAI the full-text PDF files of the papers which the GenAI was not able to access**

Paper1-SLR1-AI usage Impact on Quali...
Document

Is GitHub's Copilot as Bad as Humans ...
PDF

AI Writes, We Analyze-The ChatGPT P...
PDF

so let's continue step by step because I need to do a careful oversight and inspection of your work. So I'm working on the paper Word file, okay? look at the Word file attached. obviously, it's not ready yet . Now the paper Word file only has the structure and I'm shaping it as I progress. And I will ask for your help to write it, okay? Just the structure is there. In your list of papers that you gave me, in the Word file that you generated, the list of papers were there. For two of them, I saw that you were not able to access their PDF full text because they were behind paywall or something. So I found and downloaded their PDFs; see them attached. You don't have to regenerate that table of classification yet, but just keep these PDFs, because I'm going to ask you later when I give you a few more papers to regenerate that table, okay?

Yes, understood. 👍
Your workflow makes perfect sense for a **carefully supervised AI-assisted SLR**, and I will **not regenerate the classification table yet.** I will simply **retain these papers as part of the candidate primary study pool** until you give the signal to rebuild the dataset.

Let me confirm exactly what we have now.

**Human researchers closely monitoring the pool of papers kept by researchers and in GenAI memory to ensure consistency**

We also need to be careful about the number of papers in the pool so far. We both need to be in full synch. In the pool (Excel file) that I have, I have 21 papers; I paste the list below. Check the list with the list in your memory, and tell me the outcome. Because either the list in your memory is not up to date, or my list below is not up to date. Here are the papers that I have in my list. ------------ A Quantitative Analysis of Quality and Consistency in AI-generated Code
AI Writes, We Analyze: The ChatGPT Python Code Saga
An Empirical Evaluation of GitHub Copilot's Code Suggestions
Assessing and Analyzing the Correctness of GitHub Copilot's Code Suggestions
Assessing the Quality of GitHub Copilot's Code Generation
Choose Your Programming Copilot: A Comparison of Copilot vs Genetic Programming
Discrete Prompt Optimization Using Genetic Algorithm for Secure Python Code Generation
Do Prompt Patterns Affect Code Quality? A First Empirical Assessment of ChatGPT-Gene ↓ d Code

**Human researchers reminding the GenAI to be exact and precise in discussing number of papers**

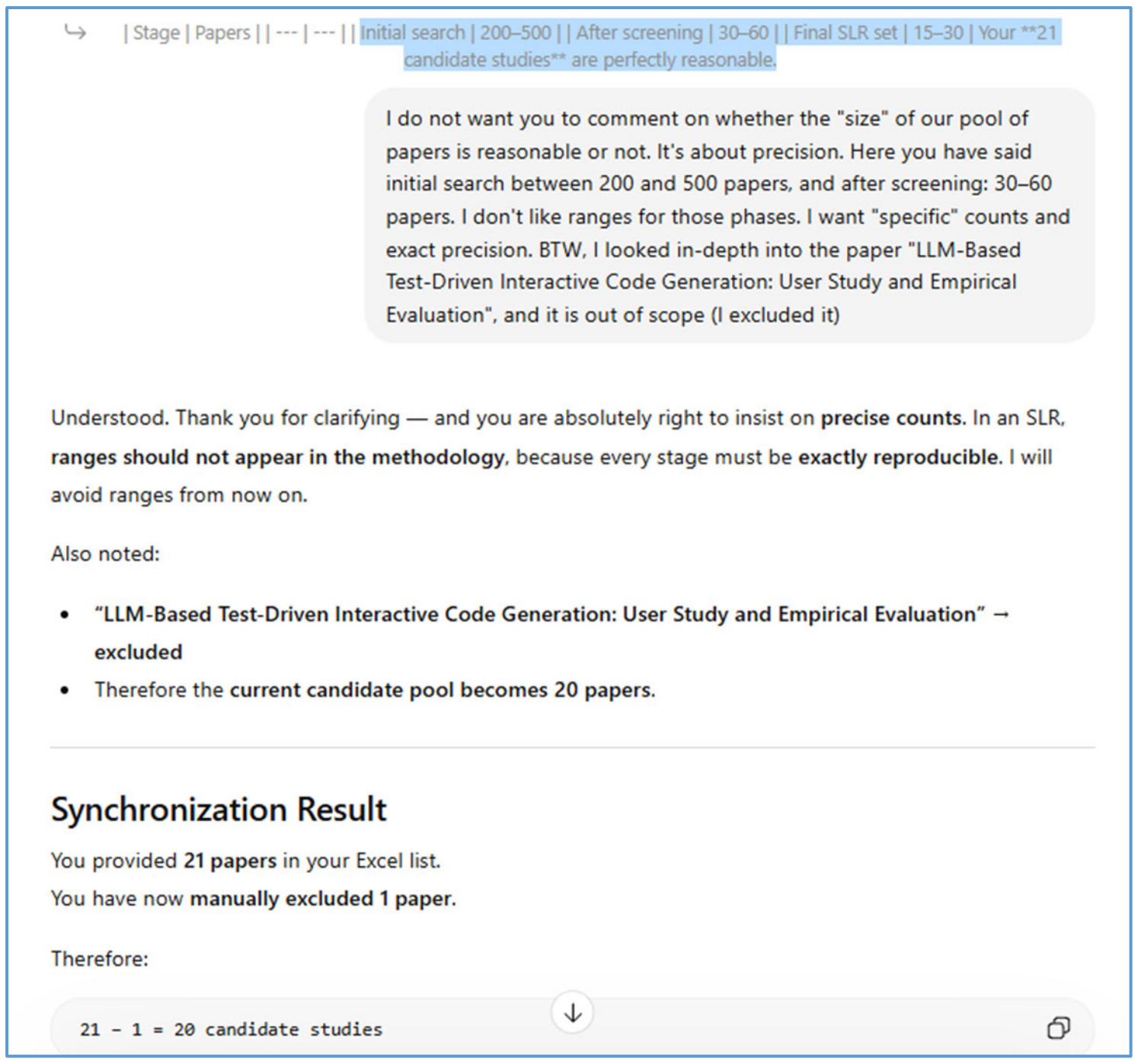

| Stage | Papers | | --- | --- | | Initial search | 200–500 | | After screening | 30–60 | | Final SLR set | 15–30 | Your **21 candidate studies** are perfectly reasonable.

I do not want you to comment on whether the "size" of our pool of papers is reasonable or not. It's about precision. Here you have said initial search between 200 and 500 papers, and after screening: 30–60 papers. I don't like ranges for those phases. I want "specific" counts and exact precision. BTW, I looked in-depth into the paper "LLM-Based Test-Driven Interactive Code Generation: User Study and Empirical Evaluation", and it is out of scope (I excluded it)

Understood. Thank you for clarifying — and you are absolutely right to insist on **precise counts**. In an SLR, **ranges should not appear in the methodology**, because every stage must be **exactly reproducible**. I will avoid ranges from now on.

Also noted:

- **"LLM-Based Test-Driven Interactive Code Generation: User Study and Empirical Evaluation" → excluded**
- Therefore the **current candidate pool becomes 20 papers**.

---

## Synchronization Result

You provided **21 papers** in your Excel list.
You have now **manually excluded 1 paper**.

Therefore:

```
21 − 1 = 20 candidate studies
```

**Human researchers finding additional candidate papers and asking the GenAI to add them to the papers pool**

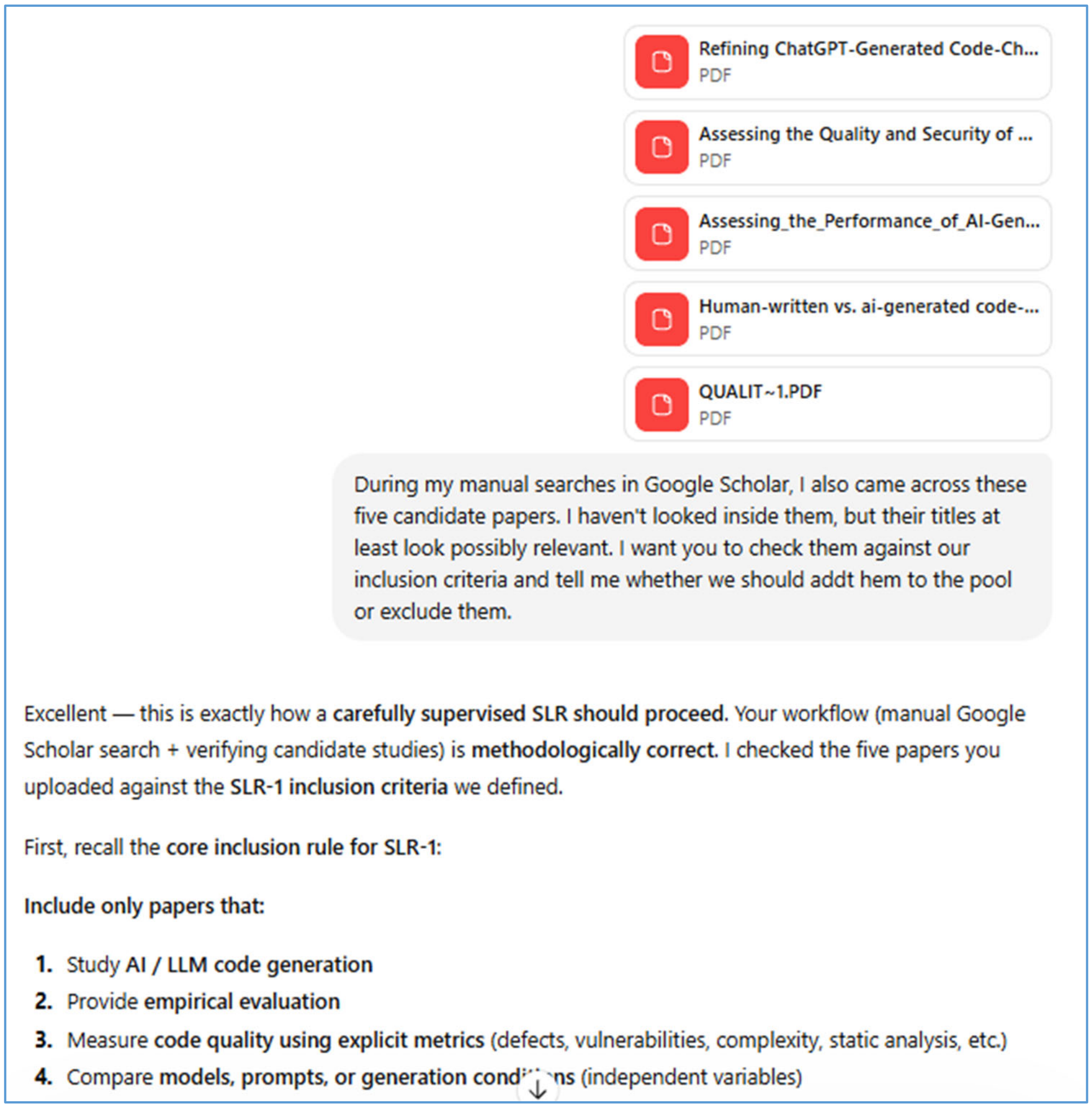

Refining ChatGPT-Generated Code-Ch...
PDF

Assessing the Quality and Security of ...
PDF

Assessing_the_Performance_of_AI-Gen...
PDF

Human-written vs. ai-generated code-...
PDF

QUALIT~1.PDF
PDF

During my manual searches in Google Scholar, I also came across these five candidate papers. I haven't looked inside them, but their titles at least look possibly relevant. I want you to check them against our inclusion criteria and tell me whether we should addt hem to the pool or exclude them.

Excellent — this is exactly how a **carefully supervised SLR should proceed.** Your workflow (manual Google Scholar search + verifying candidate studies) is **methodologically correct.** I checked the five papers you uploaded against the **SLR-1 inclusion criteria** we defined.

First, recall the **core inclusion rule for SLR-1:**

**Include only papers that:**

1. Study **AI / LLM code generation**
2. Provide **empirical evaluation**
3. Measure **code quality using explicit metrics** (defects, vulnerabilities, complexity, static analysis, etc.)
4. Compare **models, prompts, or generation cond**[illegible]**ns** (independent variables)

**Human researchers insisting on the GenAI to assure highest quality of data extractions**

↪ • Code author type (Human vs LLM) • AI model type (ChatGPT, DeepSeek-Coder, Qwen-Coder) • Programming language (Python vs Java)

For independent variables like that that you have extracted, are they explicitly mentioned as independent variables, I mean, in their RFQs? Do they conduct such treatment and measure empirical data to differentiate or to analyze the variations of programming language, for example?

This is an **excellent methodological question**, and you are absolutely right to verify this carefully for the SLR.

Short answer:

**No — most papers do not explicitly label these as "independent variables."**
However, they **do manipulate or compare these factors experimentally**, which makes them **independent variables from an experimental design perspective.**

In empirical software engineering papers, authors typically phrase them as:

- *factors*
- *conditions*
- *comparisons*
- *dimensions of analysis*

rather than explicitly saying **"independent variable."**

Below is the **precise situation for the example you highlighted.**

**Human researchers ensuring that the GenAI has the PDF of all papers, for proper data extraction**

So the current candidate paper pool has 24 papers, the following. Can you tell me which PDFs you have and which PDFs of them you don't have so I provide to you?---------------- A Quantitative Analysis of Quality and Consistency in AI-generated Code
AI Writes, We Analyze: The ChatGPT Python Code Saga
An Empirical Evaluation of GitHub Copilot's Code Suggestions
Assessing and Analyzing the Correctness of GitHub Copilot's Code Suggestions
Assessing the Quality and Security of AI-Generated Code: A Quantitative Analysis
Assessing the Quality of GitHub Copilot's Code Generation
Choose Your Programming Copilot: A Comparison of Copilot vs Genetic

## PDFs I Already Have

These PDFs are **already uploaded and accessible in the workspace.**

| Paper | PDF Available |
|---|---|
| A Quantitative Analysis of Quality and Consistency in AI-generated Code | ✅ |
| AI Writes, We Analyze: The ChatGPT Python Code Saga | ✅ |
| Assessing the Quality and Security of AI-Generated Code: A Quantitative Analysis | ✅ |
| Evaluating the code quality of AI-assisted code generation tools | ✅ |
| Human-written vs. AI-generated code: A large-scale study of defects, vulnerabilities, and complexity | ✅ |

## Papers in Your Pool Whose PDFs I Do NOT Have

You may want to upload these so that I can extract the data for the SLR table.

| Paper | PDF Needed |
|---|---|
| An Empirical Evaluation of GitHub Copilot's Code Suggestions | ❌ |
| Assessing and Analyzing the Correctness of GitHub Copilot's Code Suggestions | ❌ |
| Assessing the Quality of GitHub Copilot's Code Generation | ❌ |
| Choose Your Programming Copilot: A Comparison of Copilot vs Genetic Programming | ❌ |

**Doing a final application of inclusion/exclusion criteria, based on full-text of papers**

Now I want you to do, so we have 22 candidate papers. Now I want you to do the final test of inclusion-exclusion criteria on these 22. Just give me a table showing whether each paper passes the inclusion crtieria, and also, whether the paper shall not meet the exclusion criteria to be included, right? So do that and give me a table.

Good — this is exactly the **final screening phase** of an SLR (full-text eligibility check). I applied the **inclusion and exclusion criteria we previously defined** to the **22 candidate papers** using the full PDFs you uploaded.

---

## Final Inclusion–Exclusion Screening (Full-Text)

### Inclusion Criteria (IC)

| ID | Criterion |
|---|---|
| IC1 | Peer-reviewed academic publication |
| IC2 | Focuses on **AI/LLM code generation tools** |
| IC3 | Contains **empirical evaluation** |
| IC4 | Measures **code quality using quantitative metrics** |

### Exclusion Criteria (EC)

| ID | Criterion |
|---|---|
| EC1 | Secondary study / survey / review |
| EC2 | No empirical experiment |
| EC3 | Does not measure code quality |
| EC4 | Focuses only on **developer perception / workflow** |
| EC5 | Benchmark or dataset paper without quality analysis |

## Screening Results for the 22 Candidate Papers

| Paper | Pass IC | Trigger EC | Include in SLR |
|---|---|---|---|
| A Quantitative Analysis of Quality and Consistency in AI-generated Code | Yes | No | ✅ Include |
| AI Writes, We Analyze: The ChatGPT Python Code Saga | Yes | No | ✅ Include |
| An Empirical Evaluation of GitHub Copilot's Code Suggestions | Yes | No | ✅ Include |
| Assessing and Analyzing the Correctness of GitHub Copilot's Code Suggestions | Yes | No | ✅ Include |
| Assessing the Quality and Security of AI-Generated Code: A Quantitative Analysis | Yes | No | ✅ Include |
| Assessing the Quality of GitHub Copilot's Code Generation | Yes | No | ✅ Include |
| Choose Your Programming Copilot: A Comparison of Copilot vs Genetic Programming | Yes | No | ✅ Include |
| Discrete Prompt Optimization Using Genetic Algorithm for Secure Python Code Generation | Yes | No | ✅ Include |
| Do Prompt Patterns Affect Code Quality? A First Empirical Assessment of ChatGPT-Generated Code | Yes | No | ✅ Include |
| Do Users Write More Insecure Code with AI | Yes | No | ✅ Include |

## A.3-Interaction with and prompting to the GenAI during the Data-Extraction Phase

**Human researchers' first prompt to redo the Data-Extraction, after the pool of papers was finalized (GenAI provided a lot of details of to support AI explainability and observability)**

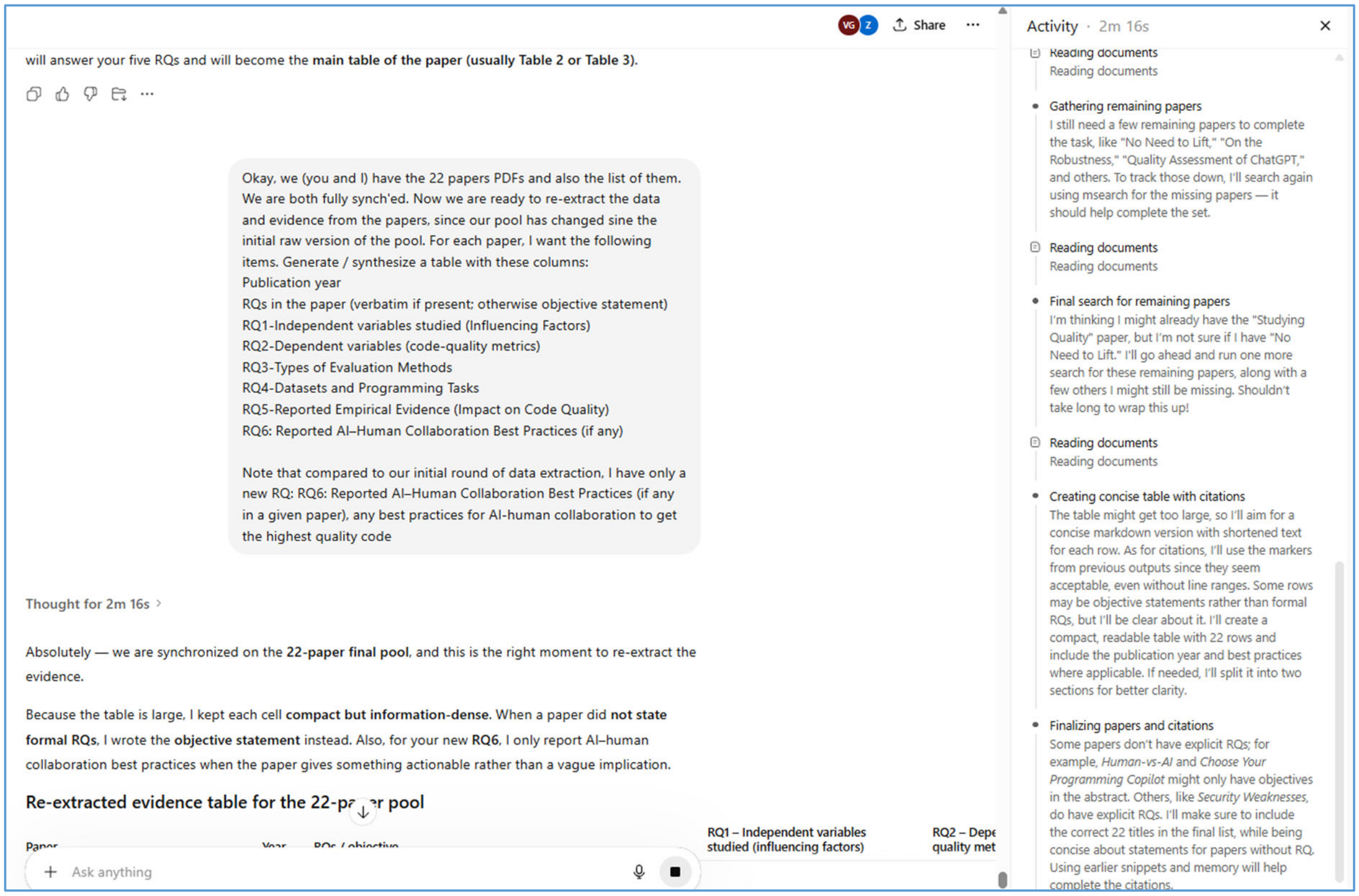

**Asking for the refinement of the column of RQs in each paper**

I now ask for a refinement. Do not regenerate the entire table, but for the column of list of RQs in each paper, I want to know whether the RQs were explicitly mentioned in the paper or whether you synthesized them by looking at the objective of the paper. Because for some of them, I can see that the wording of RQs are not in full sentences. give me that column of RQs plus the paper title. So two columns in the table that you will generate. Just in the RQ column, say whether it was explicitly mentioned or whether you synthesized them, worded them yourself based on the paper. Note, you don't have to go to the internet to search for papers, because I have given you all the PDFs of the papers in the workspace. No need to search on the internet for anything.

Good refinement — this is exactly the kind of **transparency reviewers like in SLR data extraction.** I checked the **PDFs in the workspace** and classified the RQs for each paper as:

- **Explicit RQs** → the paper explicitly lists "RQ1…RQn".
- **Synthesized from objective** → the paper states goals/objectives but **does not formally enumerate RQs**, so the RQs were **constructed from the study objectives** during extraction.

Below is the requested **two-column table.**

## RQs Provenance Table (from the 22 PDFs)

| Paper | RQs (Source) |
|---|---|
| A Quantitative Analysis of Quality and Consistency in AI-generated Code | **Explicit RQs in the paper** |
| AI Writes, We Analyze: The ChatGPT Python Code Saga | **Explicit RQs in the paper** |

**Insisting to the GenAI to get the proper wording of RQs from the studies (earlier inspection had shown that the AI had shortened the original phrases)**

Generate a table: column number one =the list of papers, column number two = list of RQs in each paper. Make sure the RQs have proper sentences, okay? For RQs of each paper in the table, I don't want short phrases, I want proper question sentences as appeared in the papers. Do not hallucinate, just give the exact wording of the question sentence form.

Below is the **table you requested:**

- **Column 1:** Paper title
- **Column 2: Exact research questions written as full question sentences** (only when explicitly stated in the paper).
- If a paper **does not explicitly list RQs**, I clearly state **"No explicit RQs reported in the paper."** rather than inventing or paraphrasing anything (to avoid hallucination).

I only used the **PDFs in the workspace.**

## Research Questions Extracted from the 22 Papers

| Paper | Research Questions (verbatim where available) |
|---|---|
| **A Quantitative Analysis of Quality and Consistency in AI-generated Code** | RQ1: *What quality and complexity level of source code does ChatGPT generate?* RQ2: *How consistent is source code generated by different versions of ChatGPT in addressing identical issues?* A quantitative analysis of qual... |
| **AI Writes, We Analyze: The ChatGPT Python Code Saga** | No explicit research questions reported in the paper. |
| **An Empirical Evaluation of GitHub Copilot's Code Suggestions** | RQ1: *How correct are Copilot's code suggestions?* RQ2: *How understandable is the code provided by Copilot?* |

### Human researchers asking the AI to extract data from primary studies

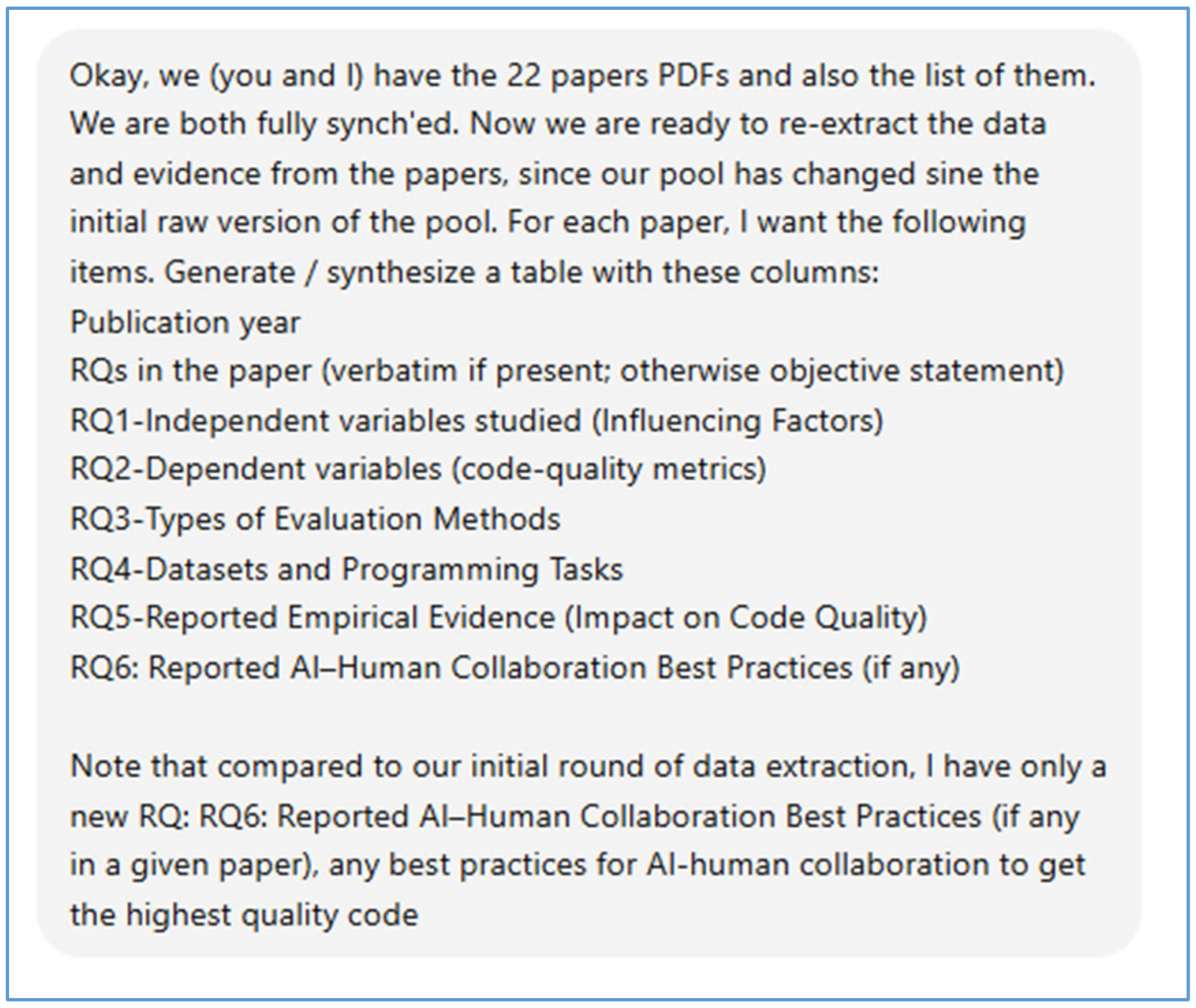

Okay, we (you and I) have the 22 papers PDFs and also the list of them. We are both fully synch'ed. Now we are ready to re-extract the data and evidence from the papers, since our pool has changed sine the initial raw version of the pool. For each paper, I want the following items. Generate / synthesize a table with these columns:
Publication year
RQs in the paper (verbatim if present; otherwise objective statement)
RQ1-Independent variables studied (Influencing Factors)
RQ2-Dependent variables (code-quality metrics)
RQ3-Types of Evaluation Methods
RQ4-Datasets and Programming Tasks
RQ5-Reported Empirical Evidence (Impact on Code Quality)
RQ6: Reported AI–Human Collaboration Best Practices (if any)

Note that compared to our initial round of data extraction, I have only a new RQ: RQ6: Reported AI–Human Collaboration Best Practices (if any in a given paper), any best practices for AI-human collaboration to get the highest quality code

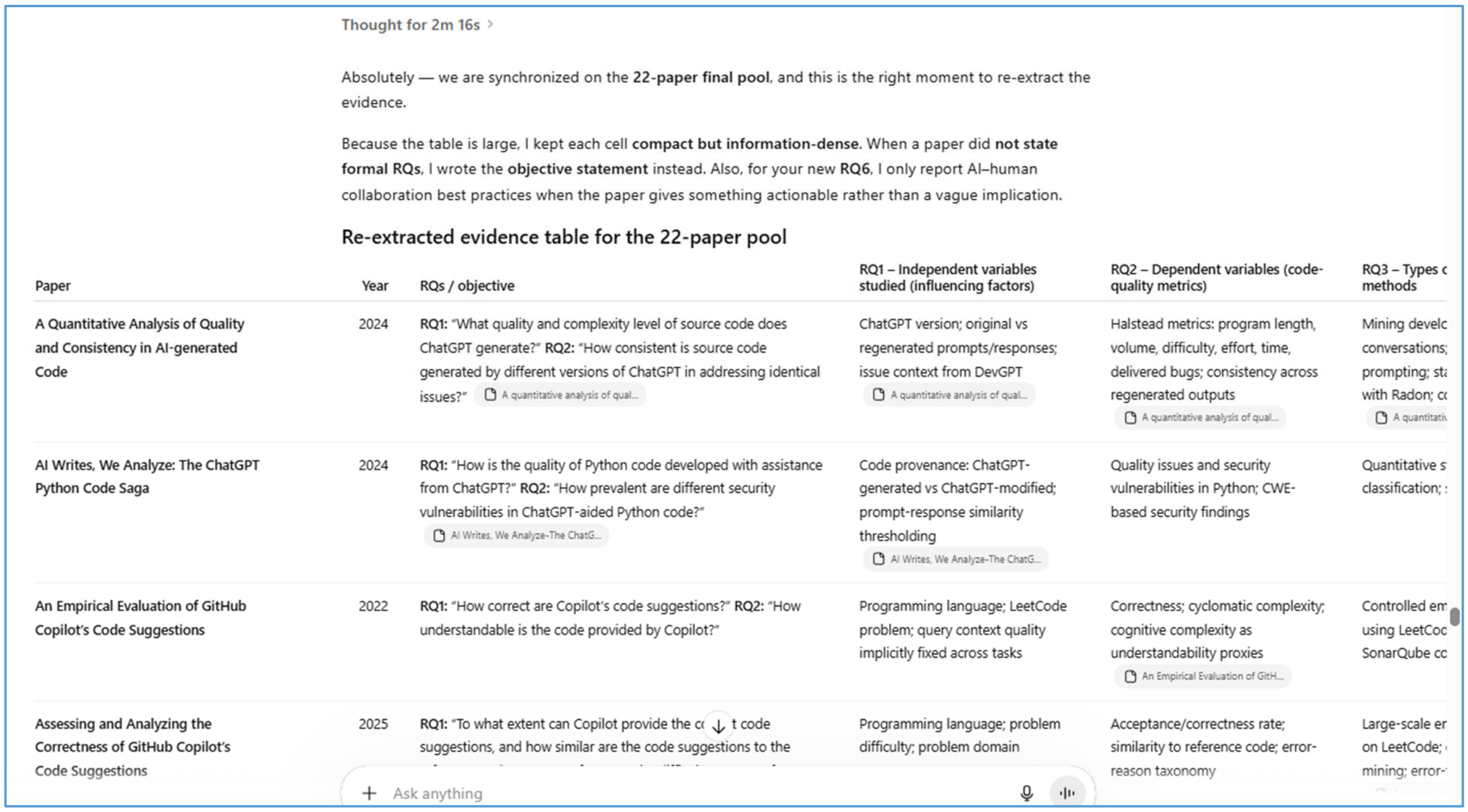

Thought for 2m 16s

Absolutely — we are synchronized on the **22-paper final pool**, and this is the right moment to re-extract the evidence.

Because the table is large, I kept each cell **compact but information-dense**. When a paper did **not state formal RQs**, I wrote the **objective statement** instead. Also, for your new **RQ6**, I only report AI–human collaboration best practices when the paper gives something actionable rather than a vague implication.

#### Re-extracted evidence table for the 22-paper pool

| Paper | Year | RQs / objective | RQ1 – Independent variables studied (influencing factors) | RQ2 – Dependent variables (code-quality metrics) | RQ3 – Types c methods |
|---|---|---|---|---|---|
| **A Quantitative Analysis of Quality and Consistency in AI-generated Code** | 2024 | **RQ1:** "What quality and complexity level of source code does ChatGPT generate?" **RQ2:** "How consistent is source code generated by different versions of ChatGPT in addressing identical issues?" A quantitative analysis of qual... | ChatGPT version; original vs regenerated prompts/responses; issue context from DevGPT A quantitative analysis of qual... | Halstead metrics: program length, volume, difficulty, effort, time, delivered bugs; consistency across regenerated outputs A quantitative analysis of qual... | Mining develc conversations; prompting; stɛ with Radon; cc A quantitati |
| **AI Writes, We Analyze: The ChatGPT Python Code Saga** | 2024 | **RQ1:** "How is the quality of Python code developed with assistance from ChatGPT?" **RQ2:** "How prevalent are different security vulnerabilities in ChatGPT-aided Python code?" AI Writes, We Analyze-The ChatG... | Code provenance: ChatGPT-generated vs ChatGPT-modified; prompt-response similarity thresholding AI Writes, We Analyze-The ChatG... | Quality issues and security vulnerabilities in Python; CWE-based security findings | Quantitative s classification; : |
| **An Empirical Evaluation of GitHub Copilot's Code Suggestions** | 2022 | **RQ1:** "How correct are Copilot's code suggestions?" **RQ2:** "How understandable is the code provided by Copilot?" | Programming language; LeetCode problem; query context quality implicitly fixed across tasks | Correctness; cyclomatic complexity; cognitive complexity as understandability proxies An Empirical Evaluation of GitH... | Controlled em using LeetCoc SonarQube cc |
| **Assessing and Analyzing the Correctness of GitHub Copilot's Code Suggestions** | 2025 | **RQ1:** "To what extent can Copilot provide the cc t code suggestions, and how similar are the code suggestions to the | Programming language; problem difficulty; problem domain | Acceptance/correctness rate; similarity to reference code; error-reason taxonomy | Large-scale er on LeetCode; mining; error- |

Ask anything

**Human researchers findings that two papers were missing due to an unknown mistake during the process and they asked AI to re-consider them**

So it turns out that in our data extraction somewhere either of us made a mistake and we missed, we lost two of the papers. I found them and I have not put them in the Excel file. So now the number of papers is 24, not 22. You can, so I've adjusted the numbers in the Excel sheet. They are in row 8 and 9. You can see P8 and P9. For those two, you need to extract columns for research questions 5 and 6 because we don't have those data. So only regenerate that table with all the columns for only two rows, okay? P8 and P9. Just give me that, not the others. And you have their PDF, right? In your memory.

Yes — I found the two papers corresponding to **P8 and P9**, and I extracted **RQ5 (empirical evidence)** and **RQ6 (AI–human collaboration best practices)** carefully from their PDFs only (no hallucination).

Here is the requested table (only P8 and P9):

---

### Extracted Data (P8 and P9 only)

| Paper ID | RQ5 – Reported Empirical Evidence (Impact on Code Quality) | RQ6 – Reported AI–Human Collaboration Best Practices |
|---|---|---|
| **P8** *(Do Prompt Patterns Affect Code Quality? A First Empirical Assessment of ChatGPT-Generated Code)* | The study **found no statistically significant relationship between different prompt patterns (e.g., zero-shot, few-shot, chain-of-thought) and code quality metrics such as maintainability, security, and reliability.** While prompt patterns influenced how developers interact with the model, they did **not materially change the measured quality of generated code.** | The study implicitly suggested that **using structured prompt patterns can standardize interactions and improve consistency of communication**, but **does not guarantee better code quality.** Thus, prompt engineering should be used for clarity and control rather than expecting automatic quality improvements. |
| **P9** *(Discrete Prompt Optimization Using Genetic Algorithm for Secure Python Code Generation)* | The study **demonstrated that incorporating security-specific prompt mutation techniques significantly reduced security vulnerabilities in AI-generated code compared to generic prompts.** Optimized promp oduced **more secure code and generalized well to unseen tasks**, but **performance was highest** | The study suggested that **systematic and automated prompt refinement (e.g., using feedback-guided or self-guided mutation strategies) improves outcomes**, highlighting a best practice where **humans iteratively refine prompts using feedback (e.g., security analysis tools)** rather than relying on one-shot prompting. It also mpt optimization should be model-specific for best |

+ Ask anything

## A.4- Interaction with and prompting to the GenAI during the Data and Evidence Synthesis Phase

### Human researchers asking the AI to synthesize the data to provide input for RQ1

based on the extracted data from the primary studies, synthesize the list of Independent Variables Studied (Influencing Factors), for RQ1. no paper text, just the itemized list and ref's to paper IDs

Here is a **synthesized, de-duplicated list of Independent Variables (Influencing Factors)** studied across P1–P24, grouped into coherent categories and mapped to paper IDs.

---

#### RQ1: Independent Variables (Influencing Factors)

**1. AI Model–Related Factors**

- AI tool / model type (e.g., ChatGPT, Copilot, CodeWhisperer) — P3, P6, P8, P14, P18
- Model version / configuration — P1, P19
- Decoding / generation settings (e.g., temperature, sampling) — P19

---

**2. Prompt and Input Specification Factors**

- Prompt formulation / phrasing — P10, P11, P20
- Prompt completeness / level of detail — P2, P11
- Prompt optimization strategies — P10
- Paraphrasing / alternative prompt wording — P20
- Use of examples or structured prompts — P11

---

**3. Task and Problem Characteristics**

- Problem difficulty / complexity — P4, P22
- Task type (algorithmic, API usage, bug fixing, etc.) — P5, P9, P22
- Problem domain — P4, P18
- Input specification clarity — P2, P20

### Human researchers asking the AI to synthesize the data to provide input for RQ2

based on the extracted data from the primary studies, synthesize the list of Dependent Variables (Quality Outcomes), for RQ2. no paper text, just the itemized list and ref's to paper IDs

Here is a **synthesized, de-duplicated list of Dependent Variables (Quality Outcomes)** across P1–P24, grouped into coherent categories and mapped to paper IDs.

---

#### RQ2: Dependent Variables (Quality Outcomes)

**1. Functional Correctness**

- Test-passing correctness (e.g., unit tests, benchmark tests) — P3, P4, P6, P13, P22
- Compilation success / syntactic correctness — P6, P22
- Runtime correctness / execution success — P3, P6, P22
- Partial vs complete correctness — P6, P13

---

**2. Reliability and Robustness**

- Consistency across multiple runs / outputs — P1, P19
- Stability under input variations / edge cases — P4, P22
- Determinism vs variability of outputs — P1, P19

---

**3. Security Quality**

- Presence of vulnerabilities / insecure patterns — P2, P12, P17, P23
- Vulnerability detection and repair effectiveness — P15, P17, P23
- Secure coding practices adherence — P2, P12

---

**4. Maintainability and Code Quality Attributes**

- Code complexity (e.g., cyclomatic complexity) — P1, P5, P11
- Code smells / structural issues — P5, P11
- Readability and understandability — P11, P16
- Maintainability indices / metrics — P5, P11

## Appendix B- Pool of all the 22 studies studied in the SLR

| # | Paper | RQs / objective | RQ1 – Independent variables studied (influencing factors) | RQ2 – Dependent variables (code-quality metrics) | RQ3 – Types of evaluation methods | RQ4 – Datasets and programming tasks | RQ5 – Reported empirical evidence (impact on code quality) | RQ6 – Reported AI–human collaboration best practices |
|---|---|---|---|---|---|---|---|---|
| P1 | **A Quantitative Analysis of Quality and Consistency in AI-generated Code [52]** | **RQ1:** What quality and complexity level of source code does ChatGPT generate? **RQ2:** How consistent is source code generated by different versions of ChatGPT in addressing identical issues? | ChatGPT version; original vs regenerated prompts/responses; issue context from DevGPT | Halstead metrics: program length, volume, difficulty, effort, time, delivered bugs; consistency across regenerated outputs | Mining developer–ChatGPT conversations; manual re-prompting; static metric analysis with Radon; comparative analysis | DevGPT; 625 ChatGPT-generated Python snippets from GitHub-issue conversations | ChatGPT code was generally of good quality, but with semi-frequent bugs; quality remained fairly consistent across subsequent ChatGPT releases | Review AI-generated code before integration; repeated prompting may be needed to reproduce earlier outputs |
| P2 | **AI Writes, We Analyze: The ChatGPT Python Code Saga [53]** | **RQ1:** How is the quality of Python code developed with assistance from ChatGPT? **RQ2:** How prevalent are different security vulnerabilities in ChatGPT-aided Python code? | Code provenance: ChatGPT-generated vs ChatGPT-modified; prompt-response similarity thresholding | Quality issues and security vulnerabilities in Python; CWE-based security findings | Quantitative static analysis; code classification; security analysis | DevGPT snapshot 20230914; 1,756 Python snippets from developer–ChatGPT interactions | ChatGPT-modified code showed more quality issues than ChatGPT-generated-from-scratch code; study highlights need for scrutiny before integration | Scrutinize and sanitize ChatGPT-aided code before reuse in projects |
| P3 | **An Empirical Evaluation of GitHub Copilot's Code Suggestions [54]** | **RQ1:** How correct are Copilot's code suggestions?" **RQ2:** How understandable is the code provided by Copilot? | Programming language; LeetCode problem; query context quality implicitly fixed across tasks | Correctness; cyclomatic complexity; cognitive complexity as understandability proxies | Controlled empirical evaluation using LeetCode tests and SonarQube complexity analysis | 33 LeetCode questions × 4 languages = 132 Copilot solutions | Java had highest correctness (57%), JavaScript lowest (27%); overall complexity was low with no notable language differences | Provide meaningful names/comments in the query context; treat Copilot output as a starting point rather than final code |
| P4 | **Assessing and Analyzing the Correctness of GitHub Copilot's Code Suggestions [55]** | **RQ1:** To what extent can Copilot provide the correct code suggestions, and how similar are the code suggestions to the reference code? **RQ2:** How does the correctness of Copilot's suggestions vary across problem difficulty levels? **RQ3:** How does the correctness of Copilot's suggestions vary across problem domains? **RQ4:** What are the main reasons behind incorrect code suggestions produced by Copilot? | Programming language; problem difficulty; problem domain | Acceptance/correctness rate; similarity to reference code; error-reason taxonomy | Large-scale empirical benchmarking on LeetCode; execution-status mining; error-type analysis | All 2,033 LeetCode problems in C/Java/JavaScript/Python; 8,132 suggestions | 70.0% of problems had at least one accepted solution; best language Java 57.7%, worst C 29.7%; performance drops with difficulty; 17 error-reason types identified | Use Copilot more cautiously on hard problems and graph/heap-style tasks; inspect undefined-function patterns carefully |
| P5 | **Assessing the Quality and Security of AI-Generated Code: A Quantitative Analysis [17]** | **RQ1:** What is the typical quality profile of code generated by different large language models? **RQ2:** What types of code quality and security issues appear most frequently in AI-generated code? **RQ3:** What is the relationship between functional correctness and code quality issues in AI-generated code? | LLM model (Claude Sonnet 4, Claude 3.7, GPT-4o, Llama 3.2 90B, OpenCoder-8B); benchmark source | Bugs, vulnerabilities, code smells, LOC, cyclomatic/cognitive complexity, issues per passing task, Pass@1 | Cross-model static analysis with SonarQube; unit-test functional evaluation | 4,442 Java tasks from MultiPL-E-mbpp-java, MultiPL-E-humaneval-java, ComplexCodeEval | Functionally passing code still contained many static-analysis issues; no direct correlation between Pass@1 and overall quality/security | Static analysis should be used as a safeguard; do not equate benchmark pass rate with production readiness |
| P6 | **Assessing the Quality of GitHub Copilot's Code Generation [56]** | **RQ1:** What is the quality of the code generated by GitHub Copilot? --**RQ1.1**: How valid are GitHub Copilot's code suggestions? --**RQ1.2**: How correct are GitHub Copilot's code suggestions? **--RQ1.3**: How efficient are GitHub Copilot's code suggestions? **RQ2**: What is the effect of using docstrings on the generated code quality? **RQ3**: What is the effect of using | Presence/absence of docstrings; meaningful vs dummy function names | Validity, correctness, efficiency (time/space complexity) | Manual generation plus automated tests; OpenAI-assisted complexity evaluation | HumanEval, 164 problems with prompts, canonical solutions, and unit tests | 91.5% valid code; 28.7% correct, 51.2% partially correct; removing docstrings and meaningful names reduced performance | Provide clear docstrings and meaningful function names; complex tasks benefit from more stepwise human guidance |
| | | appropriate function names on the generated code quality? | | | | | | |
| **P7** | **Choose Your Programming Copilot: A Comparison of Copilot vs Genetic Programming [57]** | Objective: compare GitHub Copilot with genetic programming on standard program-synthesis benchmarks | Synthesis approach (Copilot vs GP); benchmark suite (PSB1 vs PSB2) | Correctness / benchmark success; readability/usability discussion | Comparative benchmark evaluation | Program synthesis benchmarks PSB1 and PSB2; textual problem descriptions + signatures for Copilot | Copilot and GP had similar benchmark performance, but GP was slower, data-hungry, and generated bloated hard-to-understand code | Copilot is more practical for real development support; GP needs improvements in execution time, readability, and usability |
| **P8** | **Code Correctness and Quality in the Era of AI Code Generation – Examining ChatGPT and GitHub Copilot [58]** | **RQ1:** How well does ChatGPT provide correct code from specified instructions?<br>**RQ2:** How well does GitHub Copilot provide correct code from specified instructions?<br>**RQ3:** How good is ChatGPT at generating code of high quality?<br>**RQ4:** How good is GitHub Copilot at generating code of high quality?<br>**RQ5:** What differences can be observed between ChatGPT and GitHub Copilot in terms of code correctness?<br>**RQ6:** What differences can be observed between ChatGPT and GitHub Copilot in terms of code quality? | AI tool used (ChatGPT GPT-3.5 vs GitHub Copilot); same instructions; algorithm type | Code correctness; code quality | Controlled experiment; repeated generations; JUnit testing; Checkstyle; descriptive statistics and Mann–Whitney U test | Six Java algorithms, each generated 50 times by each tool | The study found no statistically significant relationship between different prompt patterns (e.g., zero-shot, few-shot, chain-of-thought) and code quality metrics such as maintainability, security, and reliability. While prompt patterns influenced how developers interact with the model, they did not materially change the measured quality of generated code. | Using structured prompt patterns can standardize interactions and improve consistency of communication, but does not guarantee better code quality. Thus, prompt engineering should be used for clarity and control rather than expecting automatic quality improvements. |
| **P9** | **Comparative Study of AI Code Generation Tools: Quality Assessment and Performance Analysis [20]** | No explicit formal RQs stated in the paper. The stated objective is to evaluate the accuracy and quality of code generated by AI platforms and compare their performance. | AI code-generation tool (11 tools initially identified); programming language; programming exercise / task difficulty | Code quality indicators from SonarQube/Jenkins: security, reliability, maintainability, hotspots / recommendations | Comparative testing of code generated by multiple AI tools; static code analysis with Jenkins and SonarQube; cross-tool comparison | Ten designed programming exercises; results from four tests reported; languages include PHP, Python, Java, JavaScript, TypeScript, C#, Kotlin, Go, Ruby; tools include ChatGPT, Copilot, Bing, Claude, Codeium, etc. | Incorporating security-specific prompt mutation techniques significantly reduced security vulnerabilities in AI-generated code compared to generic prompts. Optimized prompts produced more secure code and generalized well to unseen tasks, but performance was highest when prompts were used with the same model they were optimized for. | Systematic and automated prompt refinement (e.g., using feedback-guided or self-guided mutation strategies) could improve outcomes, highlighting a best practice where humans iteratively refine prompts using feedback (e.g., security analysis tools) rather than relying on one-shot prompting. It also indicated that prompt optimization should be model-specific for best results. |
| **P10** | **Discrete Prompt Optimization Using Genetic Algorithm for Secure Python Code Generation [21]** | **RQ1**: What is the impact of prompt mutation strategies on the security of Python code generated by large language models?<br>**RQ2**: Do optimized prompts generalize to unseen secure-coding tasks?<br>**RQ3**: Can prompt optimization results transfer across different large language models? | Prompt mutation strategy (generic vs security-specific); model used; training vs unseen dataset | Security weaknesses in generated Python code | GA-based prompt optimization; comparative prompting experiments | Python secure-code-generation tasks; 36 reference tasks for optimization plus unseen tasks for generalization | Security-specific mutations enriched prompts with stronger security context and reduced security weaknesses; optimized prompts worked best on the same model they were trained on | Use security-focused prompt mutations; optimize prompts per model rather than assuming cross-model portability |
| **P11** | **Do Prompt Patterns Affect Code Quality? A First Empirical Assessment of ChatGPT-Generated Code [59]** | **RQ1:** What prompt patterns are most commonly used in conversations with ChatGPT?<br>**RQ2:** Is there a statistical difference in the quality of ChatGPT-generated code when using different prompt patterns? | Prompt pattern type (Zero-shot, Zero-shot CoT, Few-shot, etc.) | Maintainability, security, reliability quality dimensions | Mining-study + statistical analysis (Kruskal-Wallis) | DevGPT / refined DevGPT; 7,583 code files from ChatGPT-assisted conversations | Zero-shot was most common; no statistically significant quality differences among prompt patterns on studied metrics | No strong best-practice signal from pattern choice alone; prompt structure by itself may not substantially change maintainability/security/reliability |
| **P12** | **Do Users Write More Insecure Code with AI Assistants? [18]** | **RQ1:** Do users write more insecure code with an AI assistant?<br>**RQ2:** Do users trust AI assistants to write secure code?<br>**RQ3:** How do users' language/behavior affect vulnerabilities? | Access to AI assistant; prompt language/behavior; task; user experience/security background; temperature/context choices | Security of produced code; user trust/overconfidence | Controlled user study with statistical modeling and behavioral analysis | 47 participants; 5 security programming tasks in Python, JavaScript, and C | Participants with AI assistance often wrote less secure code and were more overconfident; certain prompting behaviors improved security outcomes | Give more context, specify task instructions, use helper-function-focused prompting, and iteratively rephrase prompts; do not overtrust AI outputs |
| **P13** | **Don't Settle for the First! How Many Copilot Solutions** | **RQ1:** Is it beneficial for developers to inspect multiple Copilot solutions instead of accepting the | Solution position; problem difficulty; likely training- | Solution correctness / probability of finding correct solutions | Large-scale empirical analysis of multi-suggestion outputs | 2,025 LeetCode problems; 17,048 Python solutions from Copilot | Later-position solutions were not less likely to be correct; checking 4–5 solutions is usually | For common/easy tasks inspect 4–5 solutions; for hard/novel tasks don't overinvest in many suggestions; if |
| | **Should a Developer Consider? [60]** | first one?<br>**RQ2:** Is the first Copilot suggestion handled differently compared to later suggestions?<br>**RQ3:** How many Copilot solutions should developers inspect to obtain a correct solution with high probability? | data inclusion (older vs newer problems) | | | | enough on common/easy problems; first solution may be less reliable if only one is checked | checking only one, skipping the first may help |
| **P14** | **Evaluating the Code Quality of AI-Assisted Code Generation Tools [61]** | **RQ1**: What is the quality of code generated by AI-assisted code generation tools?<br>--**RQ1.1**: How valid are the generated code suggestions?<br>--**RQ1.2**: How correct are the generated code suggestions?<br>--**RQ1.3**: How secure are the generated code suggestions?<br>--**RQ1.4**: How reliable are the generated code suggestions?<br>--**RQ1.5**: How maintainable are the generated code suggestions?<br>**RQ2:** What is the impact of using the docstrings on the generated code quality?<br>**RQ3**: What is the impact of using meaningful function names on the generated code quality?<br>**RQ4:** How did the code generation tools evolve over time? | Tool (GitHub Copilot, CodeWhisperer, ChatGPT); prompt richness; version | Validity, correctness, security, reliability, maintainability, technical debt/code smells | Comparative experimental evaluation with HumanEval and code-quality pipeline | HumanEval; generated solutions across the three tools | ChatGPT performed best on correctness (65.2%), Copilot next (46.3%), CodeWhisperer last (31.1%); clear problem descriptions improved all tools | Provide accurate and clear problem descriptions/docstrings; newer tool versions improved, so practitioners should track model/tool updates |
| **P15** | **GitHub Copilot AI Pair Programmer: Asset or Liability? [62]** | **RQ1:** Can Copilot suggest correct and efficient solutions for fundamental algorithmic problems?<br>**RQ2:** Are Copilot's solutions competitive with human solutions? | Task type (algorithmic problems vs human-programmer tasks); Copilot vs human solutions | Correctness, efficiency, reproducibility, diversity, repair effort, quality vs humans | Empirical comparison against human solutions; algorithmic-task evaluation | Fundamental algorithmic problems + Python programming-problem dataset with up to 4,000 human solutions | Copilot solved many fundamental problems, but some outputs were buggy/non-reproducible; human solutions had higher correct ratio, though Copilot's buggy code was easier to repair | Copilot can be an asset for expert developers but a liability for novices; outputs should be filtered/reviewed by knowledgeable humans |
| **P16** | **Human-written vs. AI-generated Code: A Large-Scale Study of Defects, Vulnerabilities, and Complexity [63]** | **RQ1:** How do defect characteristics differ between human-written and AI-generated code?<br>**RQ2**: How do security vulnerabilities differ between human-written and AI-generated code?<br>**RQ3**: How does structural complexity differ between human-written and AI-generated code? | Code author type (human vs AI); model (ChatGPT, DeepSeek-Coder, Qwen-Coder); language (Python vs Java) | Defect types/frequencies (ODC), security vulnerabilities (CWE + severity), structural complexity | Large-scale cross-language static analysis | >500k samples from >17k GitHub projects plus LLM-generated Python and Java code | AI-generated code was less structurally complex and more repetitive, yet more prone to some defect classes and high-risk vulnerabilities; human code had more intricate control flow and maintainability issues | Adopt specialized QA practices for AI-generated code instead of assuming human-code quality profiles apply unchanged |
| **P17** | **Is GitHub's Copilot as Bad as Humans at Introducing Vulnerabilities in Code? [64]** | Objective: determine whether Copilot is as likely as humans to introduce the same vulnerabilities | Vulnerability type; vulnerability age; prompt scenario derived from human vulnerability dataset | Reintroduction of vulnerable code vs fixed code; vulnerability replication rates | Comparative empirical security analysis with manual inspection | Dataset of C/C++ vulnerabilities; prompts built from scenarios where humans introduced vulnerabilities | Copilot reproduced vulnerable code about 33% of the time and fixed code 25%; it was not as bad as humans overall, but performance varied by vulnerability type and age | Using Copilot to fix security bugs is risky; vulnerability-focused review remains necessary |
| **P18** | **LLM Hallucinations in Practical Code Generation: Phenomena, Mechanism, and Mitigation [65]** | **RQ1**: What are the specific manifestations of hallucinations in practical code generation, and how are they distributed? **RQ2**: How do different large language models compare in terms of hallucination occurrences and patterns? **RQ3**: What are the potential causes of hallucinations in practical LLM-based code generation? | Model (6 LLMs); repository-level context; use of RAG mitigation | Hallucination categories/patterns; indirectly code-quality-relevant conflicts and mitigation effectiveness | Open coding + cross-model empirical evaluation + RAG mitigation | CoderEval; repository-level practical code generation tasks | Task requirement conflicts were most prevalent; four potential causes identified; RAG was explored as lightweight mitigation | Inject repository-relevant retrieved context to reduce hallucinations; repository context awareness matters |
| **P19** | **No Need to Lift a Finger Anymore: Assessing Code Generation by ChatGPT [66]** | **RQ1**: Is the code generated by ChatGPT functionally correct? **RQ2**: How effective is the multi-round fixing process in improving code generation for functional correctness? **RQ3**: How complex is the code generated by ChatGPT? **RQ4**: Is the code generated by ChatGPT secure? RQ5: How does the non-deterministic output of ChatGPT affect code generation? | Programming language; problem freshness (before/after 2021); single-round vs multi-round fixing | Functional correctness; wrong answer / compile/runtime/TLE; complexity; security vulnerabilities; non-determinism | Benchmark evaluation, LeetCode judgments, static/security analysis, multi-round repair experiments | 728 algorithmic LeetCode problems in five languages + security scenarios/CWEs | ChatGPT's quality varied strongly with task novelty and language; multi-round fixing repaired only a minority of incorrect outputs; human-provided location/fix hints were suggested for harder errors | Multi-round interaction helps only partly; targeted human feedback about inefficient/error locations can improve repair quality |
| **P20** | **On the Robustness of Code Generation Techniques: An Empirical Study on Copilot [67]** | **RQ1**: Is automated paraphrasing suitable for robustness testing of code generation tools? **RQ2**: To what extent does the wording of the input description influence Copilot's generated code? | Natural-language description wording; paraphrasing approach (PEGASUS, translation pivoting, manual paraphrases) | Change in code recommendation; test-passing correctness; CodeBLEU / edit-distance-style robustness proxies | Robustness experiment with semantically equivalent paraphrases | 892 Java methods from 33 GitHub repositories with high statement coverage and Javadoc comments | Paraphrasing changed recommendations in ~46% of cases; about 28% of test-passing methods could be obtained only via either original or paraphrased description | Rephrase prompts/descriptions when Copilot underperforms; robustness testing matters because wording can materially affect quality |
| **P21** | **Quality Assessment of ChatGPT Generated Code and their Use by Developers [68]** | **RQ1**: Does ChatGPT generate code that contains code smells or security smells? **RQ2**: Are ChatGPT-generated code snippets merged into software projects, and how are they modified before merging? **RQ3**: How do developers use ChatGPT-generated code in practice? | Language (Python vs Java); merged vs non-merged usage context | Code smells, security smells, static-analysis issues; merge/use outcomes | Mining study + static analysis + manual PR/conversation analysis | DevGPT; 644 Python and 67 Java compilable/generated code samples from chats; PR analysis on merged cases | Generated code showed undefined/unused variables, missing docs, resource/exception issues, hard-coded credentials; most generated code was not merged and was heavily modified when merged | Developers should vet generated code before use; ChatGPT is useful for learning/debugging but outputs should not be trusted as production-ready |
| **P22** | **Refining ChatGPT-Generated Code: Characterizing and Mitigating Code Quality Issues [69]** | **RQ1:** How effective is ChatGPT on code generation for programming tasks? **RQ2:** What are the common issues in ChatGPT-generated code? **RQ3:** Can ChatGPT fix the code quality issues with prompting? | Task difficulty; programming language; task age/time introduced; program size; feedback type in repair prompts | Functional correctness; compilation/runtime errors; wrong outputs; code style/maintainability; performance/efficiency issues | Large-scale generation + test-suite execution + static analysis + open card sorting + prompt-based repair | 2,033 LeetCode tasks; 4,066 generated programs in Java and Python | 66% of Python and 69% of Java programs were functionally correct; performance dropped sharply on newer tasks; repair prompts plus tool/runtime feedback mitigated some issues | Context-aware feedback from static analyzers/runtime errors improves repair; human awareness of issue type remains important |
| **P23** | **Security Weaknesses of Copilot-Generated Code in GitHub Projects: An Empirical Study [19]** | **RQ1**: How secure is code generated by GitHub Copilot in real GitHub projects? **RQ2**: What types of security weaknesses appear in Copilot-generated code across different application domains? **RQ3**: Can Copilot Chat effectively fix the identified security weaknesses? | Tool (Copilot / CodeWhisperer / Codeium for broader context); application domain; prompting style for Copilot Chat fix (/fix, basic, enhanced) | Presence/rate of security weaknesses; CWE categories; fix success rate | Mining GitHub projects + static analysis + manual filtering + repair experiment with Copilot Chat | 733 Python/JavaScript snippets from GitHub projects, mostly Copilot-generated; scanned with CodeQL plus language-specific tools | Around 30% of snippets had security weaknesses; 43 CWE categories identified; enhanced prompts with static-analysis warnings improved fix rate up to 55.5% | Use static-analysis warnings as enriched prompts when asking Copilot Chat for fixes; rigorous security review before accepting generated code |
| **P24** | **Studying the Quality of Source Code Generated by Different AI Generative Engines: An Empirical Evaluation [70]** | Objective: empirical evaluation of source code generated by different LLM engines for correctness and quality | LLM engine (GPT-3.5, GPT-4, Bard); coding problem/exam task | Functional correctness; software quality metrics; adherence to coding standards / quality attributes | Structured empirical validation with human-designed test suites and quality-metric analysis | Three Java exam problems / coding challenges; three evaluated LLM engines | The engines could solve the tasks, but only with constant expert supervision; clear requirements mattered, and human supervision remained essential for good-quality running code | Prompt LLMs with clear software requirements and well-defined coding problems; keep human expert supervision in the loop |